\documentstyle[12pt,fleqn,here,epsf,epsfig,cite]{article}
\renewcommand{\theequation}{\arabic{equation}}

\def\lsim{\raise0.3ex\hbox{$\;<$\kern-0.75em\raise-1.1ex\hbox{$\sim\;$}}}
\def\gsim{\raise0.3ex\hbox{$\;>$\kern-0.75em\raise-1.1ex\hbox{$\sim\;$}}}

\newcommand{\be}[1]{\begin{equation} \label{(#1)}}
\newcommand{\ee}{\end{equation}}
\newcommand{\baq}[1]{\begin{eqnarray} \label{(#1)}}
\newcommand{\eaq}{\end{eqnarray}}
\newcommand{\ba}{\begin{array}}
\newcommand{\ea}{\end{array}}

\begin{document}
\setlength{\unitlength}{1cm}
\setlength{\mathindent}{0cm}
\thispagestyle{empty}
\hfill UWThPh-2007-16\\
\null
\hfill BONN-TH-2007-02\\
\null
\hfill arXiv:0706.3822 [hep-ph]\\
\vskip .8cm
\begin{center}
{\Large \bf 
CP asymmetries with Longitudinal and Transverse Beam Polarizations
in Neutralino Production and Decay into the $Z^0$ Boson at the ILC
\\[5ex]
}
\vskip 2.5em
{\large
{\sc A.~Bartl$^{a}$,
     K.~Hohenwarter-Sodek$^{a}$,
     T.~Kernreiter$^{a}$,
     O.~Kittel$^{b}$
}}\\[2ex]
{\normalsize \it
$^{a}$ Faculty of Physics, University of Vienna, 
Boltzmanngasse 5, A-1090 Wien, Austria}\\
{\normalsize \it
$^{b}$ Physikalisches Institut der Universit\"at Bonn, 
Nu{\ss}allee 12, D-53115 Bonn, Germany}\\
%}
\vskip 1em
\end{center} \par
\vskip 1.8cm
%\vfil

\begin{abstract}
We study neutralino production 
at the linear collider with 
the subsequent two-body decays $\tilde\chi^0_i \to \tilde\chi^0_n Z^0$ and 
$Z^0 \to \ell \bar\ell$, with $\ell=e,\mu,\tau$, or $Z^0 \to q\bar q$ with $q=c,b$.
We show that transverse electron and positron beam polarizations allow the 
definition of unique CP observables. These are azimuthal asymmetries in the 
distributions of the final leptons or quarks.
We calculate these CP asymmetries and the cross sections in the Minimal 
Supersymmetric Standard Model with complex higgsino and gaugino 
parameters $\mu$ and $M_1$.  For final quark pairs,
we find CP asymmetries as large as $30\%$.
We discuss the significances for observing 
the CP asymmetries at the International Linear Collider~(ILC). 
Finally we compare the CP asymmetries with those asymmetries which
require unpolarized and/or longitudinally polarized beams only.
\end{abstract}

\newpage

\section{Introduction}

Supersymmetric (SUSY) models predict new particles with
masses of the order of a few hundred GeV~\cite{haberkane,Drees:2004jm}. 
Their discovery
is a major goal of present and future colliders in the TeV range.
In particular, the International $e^+e^-$ Linear Collider
(ILC)~\cite{Aguilar-Saavedra:2001rg,Abe:2001nn,Abe:2001gc,
Weiglein:2004hn,Aguilar-Saavedra:2005pw}, 
with a center-of-mass energy of $\sqrt s=500$~GeV and an 
integrated luminosity of ${\mathcal L}=500$~fb$^{-1}$
in the first stage, will precisely measure 
the masses and couplings of the SUSY particles when they are kinematically accessible.
It has been shown that the underlying parameters of the SUSY model
can be determined at the percent level and 
better~\cite{Weiglein:2004hn,Aguilar-Saavedra:2005pw,Lafaye:2004cn,
Bechtle:2005vt}.
In particular, the option of polarized $e^+$
and $e^-$ beams~\cite{Moortgat-Pick:2005cw} at the ILC can yield 
higher statistics
to test models beyond the Standard Model (SM). Transversely polarized 
beams allow to study additional observables which 
are sensitive to effects of new physics.
These are, for example, models with extra spacial dimensions,
specific triple-gauge boson couplings,
and also new sources of CP violation~\cite{Moortgat-Pick:2005cw}.

\medskip

In the Minimal Supersymmetric Standard Model 
(MSSM)~\cite{haberkane,Drees:2004jm}, 
the spin-half superpartners of the neutral gauge 
and CP-even Higgs bosons mix and form the four neutralinos 
$\tilde\chi^0_i$.
At tree-level, the neutralino sector of the MSSM is defined by the  
$U(1)_Y$ and $SU(2)_L$ gaugino mass parameters $M_1$ and $M_2$,
respectively, the higgsino mass parameter $\mu$, and the ratio 
$\tan\beta =v_2/v_1$ of the vacuum expectation values of the two 
neutral Higgs fields.
Besides the sleptons, the superpartners of the leptons,
and the charginos, the superpartners 
of the charged gauge and Higgs bosons, the neutralinos are expected 
to be among the lightest SUSY particles in many models.
The neutralinos will be 
pair-produced~\cite{Bartl:1986hp,Moortgat-Pick:1999di,Choi:1999cc} at the ILC
\begin{eqnarray} \label{production}
        e^+e^-&\to&\tilde\chi^0_i  \tilde\chi^0_j~, 
        \quad i,j =1,\dots,4~.  
\end{eqnarray}
By measurements of the neutralino masses, cross sections and
decay distributions, methods have been developed to
determine the parameters of the neutralino 
sector~\cite{Kneur:1999nx,Barger:1999tn,Choi:2001ww,Choi:2003hm,
Choi:2005gt,Gounaris:2002pj}.
If CP is violated, the parameters
$M_1=|M_1|~e^{i\phi_{M_1}}$ and $\mu=|\mu|~e^{i\phi_{\mu}}$
can be complex, while $M_2$ and $\tan\beta$ can then be chosen 
real and positive.

\medskip

In general large values for CP phases in SUSY models
lead to theoretical predictions for the electric dipole moments (EDM)
of electron, neutron and that of the atoms $^{199}$Hg and $^{205}$Tl,
which are close or beyond the current experimental upper 
bounds~\cite{Yao:2006px,Ibrahim:2002ry,Abel:2005er}. 
The restrictions on the phases from
EDM measurements, however, strongly depend on the SUSY model, 
see e.g.~\cite{Barger:2001nu,Bartl:2003ju},
and on the scenario~\cite{Brhlik:1998zn,YaserAyazi:2006zw}. 
This means that SUSY CP phases of the order one are not ruled out
by the present EDM experiments.
For an unambiguous determination of the CP phases 
independent measurements are necessary,
e.g., by analyzing CP sensitive observables at colliders,
in particular at the ILC.

\medskip

CP asymmetries with triple products~\cite{Kizukuri:1993vh,valencia} 
have been analyzed for the production of neutralinos and for their various
two-body~\cite{Kittel:2004rp,Bartl:2003tr,Bartl:2003gr,Choi:2003pq,
Bartl:2004ut,Bartl:2005uh} and three-body 
decays~\cite{Choi:2005gt,Aguilar-Saavedra:2004dz,Kizukuri:1990iy,Bartl:2004jj}.
It has been pointed out that longitudinally polarized beams can enhance 
simultaneously the cross sections and the CP asymmetries, such that higher 
statistics allows to measure even small phases~\cite{Moortgat-Pick:2005cw}.

\medskip

In contrast to longitudinally polarized beams, the possibility of transverse
beam polarization allows us to define a whole class of new and
unique CP sensitive observables. These are asymmetries in the \emph{azimuthal} 
distributions of final state particles. In general the transverse polarization
of both beams is required, since
these asymmetries vanish if only one beam is polarized.
CP observables for transversely polarized beams have been studied
for the production of selectrons~\cite{Bartl:2006bn} and
charginos~\cite{Choi:2000ta,Bartl:2004xy}.
For neutralino production, two-body decays have been analyzed:
$\tilde\chi^0_j\to\tilde\ell^\pm_{L,R} \ell^\mp 
\to\ell^+\ell^-\tilde\chi^0_1$~($\ell=e,\mu$) 
\cite{Bartl:2005uh,Choi:2006vh},
$\tilde\chi^0_j\to \tilde\chi^0_1h^0$ and 
$\tilde\chi^0_j\to\tilde\chi^0_1Z^0$ 
\cite{Choi:2006vh}.

\medskip

In this work we study CP asymmetries with triple
products in neutralino production~(\ref{production}),
followed by the decay of one of the neutralinos into the $Z^0$ boson
\begin{eqnarray} \label{decay_1A}
\tilde\chi^0_j \to \tilde\chi^0_n  Z^0~, \quad n=1 \quad\mbox{and} 
\quad j=2,3~,
\end{eqnarray}
for longitudinal and transverse beam polarizations.
Due to the Majorana properties of the neutralinos,
the angular distribution of the $Z^0$ boson
is independent of the $\tilde\chi^0_j$ polarization, if the $Z^0$ polarization is 
summed~\cite{Bartl:2003ck,Kittel:2004rp,Bartl:2004ut,Choi:2003fs,Choi:2006vh}.
Thus for longitudinally polarized beams, all CP sensitive information is lost.
On the other hand, the CP sensitive observables in the reaction 
(\ref{production})
with transversely polarized beams
require the reconstruction of the neutralino production plane.
This can be challenging, since there is not enough 
kinematical information from energy and momentum 
conservation in the process 
$e^+ e^- \to \tilde\chi^0_1 \tilde\chi^0_1 Z^0$~\cite{Bartl:2005uh,Choi:2006vh}.
Hence in either way, the polarization of the $Z^0$ has to be 
included and analyzed by 
the angular distributions of its decay products. 
On account of this, we study its leptonic and hadronic decays  
\begin{eqnarray} \label{decay_2B}
Z^0 \to f  \bar f~, \quad  f=\ell,q~,\quad\ell=e,\mu,\tau~, \quad q=c,b~.
\end{eqnarray}
We will show that spin correlations between production and decay 
lead to sizeable CP asymmetries,
which do not require the reconstruction of the production plane.

\medskip

In Section~\ref{Definitions}, we give the Lagrangians and couplings
for neutralino production~(\ref{production}) and 
decay~(\ref{decay_1A}), (\ref{decay_2B}).
In Section~\ref{crosssection}, we present the analytical formulae for 
the amplitude squared with polarized beams.
We define the CP asymmetries for transversely and longitudinally
polarized beams in Section~\ref{asymmetry}.
In Section~\ref{Numerical results},
we present numerical results for the CP asymmetries and the cross sections.
We compare the CP asymmetries with longitudinal
and transverse beam polarizations, and show that 
transverse beam polarizations can help to 
determine the phases in the neutralino sector.
In Section~\ref{Summary and conclusion}, we give a summary 
and conclusions.

\newpage

\section{Lagrangians and couplings
  \label{Definitions}}

\begin{figure}[t]
\hspace{-2.cm}
\begin{minipage}[t]{6cm}
\begin{center}
{\setlength{\unitlength}{0.6cm}
\begin{picture}(5,5)
\put(-2.5,-8.5){\includegraphics{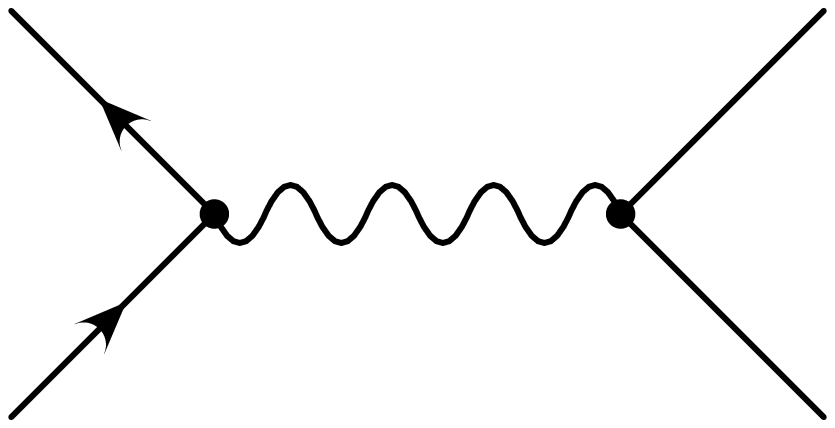}}
\put(1.7,-.4){{\small $e^{-}$}}
\put(7.8,-.4){{\small $\tilde\chi^0_j$}}
\put(1.7,3.8){{\small $e^{+}$}}
\put(7.8,3.8){{\small $\tilde\chi^0_i$}}
\put(5.1,2.5){{\small $Z^0$}}
\end{picture}}
\end{center}
\end{minipage}
\hspace{-0.5cm}
\begin{minipage}[t]{5cm}
\begin{center}
{\setlength{\unitlength}{0.6cm}
\begin{picture}(2.5,5)
\put(-4,-9){\includegraphics{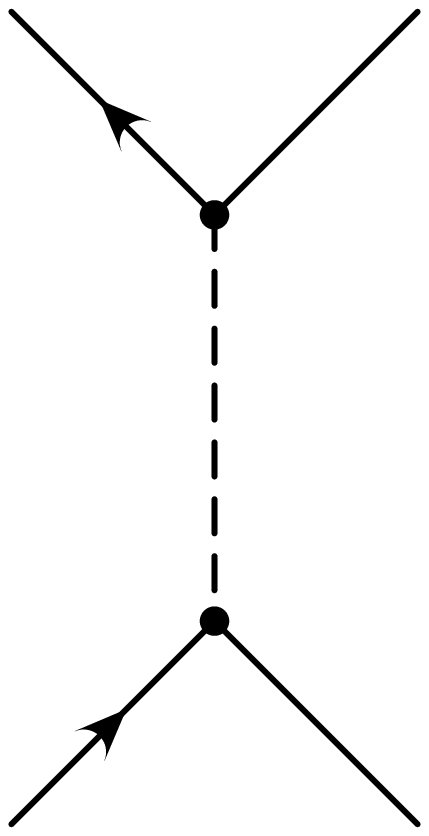}}
\put(1.8,-1.5){{\small $e^{-}$}}
\put(1.8,3.8){{\small $e^{+}$}}
\put(5.8,-1.5){{\small $\tilde\chi^0_j$}}
\put(5.8,3.8){{\small $\tilde\chi^0_i$}}
\put(4.4,1.5){{\small $\tilde{e}_{L,R}$}}
 \end{picture}}
\end{center}
\end{minipage}
\begin{minipage}[t]{5cm}
\begin{center}
{\setlength{\unitlength}{0.6cm}
\begin{picture}(2.5,5)
\put(-4.5,-9){\includegraphics{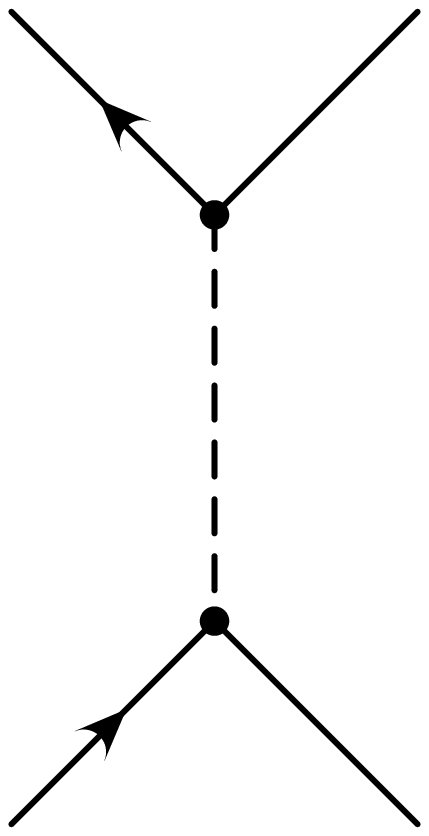}}
\put(1.3,-1.5){{\small $e^{-}$}}
\put(1.3,3.8){{\small $e^{+}$}}
\put(5.5,-1.5){{\small $\tilde\chi^0_i$}}
\put(5.5,3.8){{\small $\tilde\chi^0_j$}}
\put(3.9,1.5){{\small $\tilde{e}_{L,R}$}}
\end{picture}}
\end{center}
\end{minipage}
\vspace{.7cm}
\caption{\label{Fig:FeynProd}Feynman diagrams for neutralino production
  $e^{+}e^{-}\to\tilde{\chi}^0_i\tilde{\chi}^0_j$
\cite{Bartl:1986hp}.}
\end{figure}

In the MSSM, neutralino production~(\ref{production}) proceeds via $Z^0$ boson 
exchange in the $s$-channel, and selectron $\tilde{e}_{L,R}$ exchange in the 
$t$- and $u$-channels, see the Feynman diagrams in Fig.~\ref{Fig:FeynProd}.
The Lagrangians for neutralino production 
$e^+e^-\to\tilde\chi_i^0\tilde\chi_j^0$
and decay $\tilde\chi_j^0\to\tilde\chi_n^0 Z^0$ 
are~\cite{Bartl:1986hp,Moortgat-Pick:1999di}
\begin{eqnarray}
{\cal L}_{Z^0\tilde{\chi}^0_i\tilde{\chi}^0_j} &=&
\frac{1}{2}\frac{g}{\cos\theta_W}
Z^0_{\mu}\bar{\tilde{\chi}}^0_i\gamma^{\mu}
[O_{ij}^{''L} P_L+O_{ij}^{''R} P_R]\tilde{\chi}^0_j~, \quad i, j=1,\dots,4~, 
\label{Zchichi}\\[2mm]
{\cal L}_{e \tilde{e}\tilde{\chi}^0_i} &=&
g f^L_{e i}\bar{e}P_R\tilde{\chi}^0_i\tilde{e}_L+
g f^R_{e i}\bar{e}P_L\tilde{\chi}^0_i\tilde{e}_R+\mbox{h.c.}~, 
\label{slechie}\\[2mm]
{\cal L}_{Z^0 f \bar f} &=&-\frac{g}{\cos\theta_W}
Z^0_{\mu}\bar f\gamma^{\mu}[L_fP_L+ R_f P_R]f~,\quad f=\ell,q~,
\end{eqnarray}
with $P_{L, R}=(1\mp \gamma_5)/2$.
In the photino, zino, Higgsino basis
($\tilde{\gamma},\tilde{Z}, \tilde{H}^0_a, \tilde{H}^0_b$),
the couplings are 
\begin{eqnarray}
        &&O_{ij}^{''L}=-\frac{1}{2} \left[
        (N_{i3}N_{j3}^*-N_{i4}N_{j4}^*)\cos2\beta
  +(N_{i3}N_{j4}^*+N_{i4}N_{j3}^*)\sin2\beta \right]~,\\[2mm]
&&O_{ij}^{''R}=-O_{ij}^{''L*}~,\\[2mm]
&&f_{e i}^L = \sqrt{2}\bigg[\frac{1}{\cos
\theta_W}(\frac{1}{2}-\sin^2\theta_W)N_{i2}+\sin \theta_W
N_{i1}\bigg]~,\\[2mm]
&&f_{e i}^R = \sqrt{2} \sin \theta_W
\Big[\tan \theta_W N_{i2}^*- N_{i1}^*\Big]~,\\[2mm]
&&L_f=T_{3f}-q_{f}\sin^2\theta_W~, \quad
  R_f=-q_{f}\sin^2\theta_W \label{eq_5}~,
\label{eq_6}
\end{eqnarray}
with the weak mixing angle $\theta_W$,
the weak coupling constant $g=e/\sin\theta_W$, $e>0$,
the electric charge $q_f$ and isospin $T_{3f}$ of fermion $f$,
and the ratio $\tan \beta=v_2/v_1$ of the vacuum expectation values 
of the two Higgs fields.
The neutralino couplings $O_{ij}^{''L,R}$ and  $f^{L,R}_{e i}$ 
contain the complex mixing elements $N_{ij}$,
which diagonalize the neutralino matrix
 $N_{i \alpha}^*Y_{\alpha\beta}N_{\beta k}^{\dagger}=
 m_{\chi_i}\delta_{ik}$ \cite{haberkane}, 
with the neutralino masses $ m_{\chi_i}>0$.

\section{Distributions and cross section
     \label{crosssection}}

The CP sensitive observables for transversal and longitudinal beam polarizations are
asymmetries in the angular distributions of final state particles.
They depend on the spin correlations between  
neutralino production and decay~(\ref{production})--(\ref{decay_2B}).
The polarizations of the particles can be included
using the spin density matrix formalism~\cite{Haber:1994pe}.
In this formalism, the amplitude squared for neutralino
production has been calculated at tree level 
for unpolarized and longitudinally polarized 
beams~\cite{Moortgat-Pick:1999di}, and for 
transversely polarized beams~\cite{Bartl:2005uh}. 
The subsequent decay of one of the
neutralinos into a $Z^0$ boson, $\tilde\chi_j^0\to \tilde\chi_n^0 Z^0$, 
followed by $Z^0\to f \bar f$, has also been calculated in the
density matrix formalism~\cite{Bartl:2003ck,Bartl:2004ut,Kittel:2004rp}. 
For completeness, we shortly summarize the results in the following.

\medskip

The amplitude squared for
neutralino production~(\ref{production}) and 
decay~(\ref{decay_1A})--(\ref{decay_2B}) is given by
\begin{eqnarray} \label{amplitude2}
|T|^2 &=& 4\,|\Delta(\tilde\chi^0_j)|^2\, |\Delta(Z^0)|^2
        \left[ P~ D +  
     \sum_{b=1}^3 \Sigma_P^b~ \Sigma_D^b
    \right]~,
\end{eqnarray}
with the propagators of the decaying neutralino $\tilde\chi^0_j$ and $Z^0$ boson
\begin{equation}\label{neut:propagators}
\Delta(\tilde\chi^0_j)=\frac{i}{s_{\chi_j}-m_{\chi_j}^2
        +im_{\chi_j}\Gamma_{\chi_j}}~,\quad
\Delta(Z^0)=\frac{i}{s_Z-m^2_Z+im_Z\Gamma_Z}~,
\end{equation}
where $\Gamma_Z$ and $\Gamma_{\chi_j}$ are the total decay widths
of the associated particles and $s_Z$ and $s_{\chi_j}$
are their invariant masses.
The differential cross section then reads  
\begin{equation}\label{Z:crossection}
        d\sigma=\frac{1}{2 s}|T|^2 
        d{\rm Lips}(s;p_{\chi_j},p_{\chi_n},p_{f},p_{\bar f})~,
\end{equation}
where $d{\rm Lips}$ is the Lorentz invariant phase-space element, given 
in Appendix~\ref{Phase space}.

\medskip

The amplitude squared~(\ref{amplitude2}) has contributions
from neutralino production~($P$) and decay~($D$).
The function $P$ is independent of the polarizations
of the neutralinos, whereas
$\Sigma^b_P$ depends on the polarization
of neutralino $\tilde\chi^0_j$.
The longitudinal polarization of the neutralino is given by
$\Sigma^3_P/P$,
the transverse polarization of the neutralino in the production plane
is given by $\Sigma^1_P/P$
and the polarization perpendicular to the production plane is given by
$\Sigma^2_P/P$.
It is convenient to decompose the functions $P$ and $\Sigma^b_P$ into
unpolarized (unpol), longitudinally ($L$) and transversely polarized~($T$)
contributions of the electron and positron beams:
\begin{equation}\label{eq:defpsum}
P =P_{\rm unpol}+ P_L +P_T~, \quad
\Sigma^b_P=\Sigma^b_{P,{\rm unpol}}+ \Sigma^b_{P,L} +\Sigma^b_{P,T}~.
 \end{equation}
The quantities $P_{\rm unpol}$, $P_L$ and $\Sigma^b_{P,{\rm unpol}}$,
$\Sigma^b_{P,L}$ are given in~\cite{Moortgat-Pick:1999di}, and
the quantities $P_T$ and $\Sigma^b_{P,T}$ have been calculated
in \cite{Bartl:2005uh}.

\medskip

The quantities $\Sigma^b_{P,T}$ with transversal beam polarization 
depend on the polarization of the neutralino $\tilde\chi_j^0$ and 
are\footnote[2]{These terms are also given in \cite{Bartl:2005uh}, however, 
there in the terms $\Sigma^b_P(Z \tilde{e}_{L,R})_{T}$ and 
$P(Z \tilde{e}_{L,R})_{T}$ a factor $g^2/\cos^2\theta_W$ is missing.}
\begin{equation}
\Sigma^b_{P,T}=
\Sigma^b_P(Z \tilde{e}_L)_{T}+\Sigma^b_P(Z \tilde{e}_R)_{T}
+\Sigma^b_P(\tilde{e}_L \tilde{e}_R)_{T}~,
\label{eq:defsigbsum}
\end{equation}
where
\begin{eqnarray}
\Sigma^b_P(Z \tilde{e}_L)_{T}&=& {\cal P}_T^- {\cal P}_T^+~\frac{1}{2}
\frac{g^4}{\cos^4\theta_W}R_e\nonumber\\
&&{}\times\{{\rm Re}(
\Delta(Z)f^{L*}_{e i}f^{L}_{e j}O''^L_{ij}
[\Delta(\tilde{e}_L,u)^*-\Delta(\tilde{e}_L,t)^*])~r_1^b
\nonumber\\
&&{}
-\;\,{\rm Im}(
\Delta(Z)f^{L*}_{e i}f^{L}_{e j}O''^L_{ij}
[\Delta(\tilde{e}_L,u)^*+\Delta(\tilde{e}_L,t)^*])~r_2^b\}~,
\label{eq:SPZeL}
\end{eqnarray}
\begin{eqnarray}
\Sigma^b_P(Z \tilde{e}_R)_{T}&=& {\cal P}_T^- {\cal P}_T^+~\frac{1}{2}
\frac{g^4}{\cos^4\theta_W}L_e\nonumber\\
&&{}\times\{{\rm Re}(
\Delta(Z)f^{R*}_{e i}f^{R}_{e j}O''^R_{ij}
[\Delta(\tilde{e}_R,u)^*-\Delta(\tilde{e}_R,t)^*])~r_1^b\nonumber\\
&&{}
+\;\,{\rm Im}(
\Delta(Z)f^{R*}_{e i}f^{R}_{e j}O''^R_{ij}
[\Delta(\tilde{e}_R,u)^*+\Delta(\tilde{e}_R,t)^*])~r_2^b\}~,
\label{eq:SPZeR}
\end{eqnarray}
\begin{eqnarray}
\Sigma^b_P(\tilde{e}_L \tilde{e}_R)_{T}&=& {\cal P}_T^- {\cal P}_T^+~\frac{1}{4}
g^4 \nonumber\\
&&{}\!\!\!\!\!\!\!\!\!\!\!\!\!
\times \{
{\rm Re}([
\Delta(\tilde{e}_L,t)\Delta(\tilde{e}_R,u)^*-
\Delta(\tilde{e}_L,u)\Delta(\tilde{e}_R,t)^*]
f^{L*}_{e i}f^L_{e j}f^{R*}_{e i}f^R_{e j})~r^b_1
\nonumber\\
&&{}\!\!\!\!\!\!\!\!\!\!\!\!\!+\;\,
{\rm Im}([
\Delta(\tilde{e}_L,t)\Delta(\tilde{e}_R,u)^*+
\Delta(\tilde{e}_L,u)\Delta(\tilde{e}_R,t)^*]
f^{L*}_{e i}f^L_{e j}f^{R*}_{e i}f^R_{e j})~r^b_2\}~,
\label{eq:SPeLeR}
\end{eqnarray}
with $-1\leq {\cal P}_T^\pm \leq 1$ [$({\cal P}^\pm_T)^2+({\cal P}^\pm_L)^2\leq 1$]
being the degree of transverse $e^\pm$ beam polarizations.
The selectron propagators are 
$\Delta(\tilde{e}_{L,R},t)=
i/(t-m^2_{\tilde{e}_{L,R}})$ and
$\Delta(\tilde{e}_{L,R},u)=
i/(u-m^2_{\tilde{e}_{L,R}})$, with the Mandelstam variables 
$t=(p_{\chi_j}-p_{e^-})^2$ and $u=(p_{\chi_i}-p_{e^-})^2$.
For a center-of mass energy $\sqrt{s}$ far beyond the $Z^0$-threshold, 
the $Z^0$-width can be neglected in the propagator $\Delta(Z^0)= i/(s-m^2_Z)$.
The kinematical factors in Eqs.~(\ref{eq:SPZeL})--(\ref{eq:SPeLeR}) are
\begin{eqnarray}
 r^b_1 &=&m_{\chi_j}\{[(t_-\cdot s^b)(t_+ \cdot p_{\chi_i})+(t_-\cdot p_{\chi_i})(t_+\cdot s^b)]
(p_{e^-}\cdot p_{e^+})\nonumber\\
&& +[(p_{e^-}\cdot s^b)(p_{e^+}\cdot p_{\chi_i})+(p_{e^-}\cdot p_{\chi_i})(p_{e^+}\cdot s^b)
-(p_{e^-}\cdot p_{e^+})(p_{\chi_i}\cdot s^b)]
(t_-\cdot t_+ )\}~,
\label{eq:Sr1}
\end{eqnarray}
\begin{eqnarray}
 r^b_2&=& \varepsilon_{\mu\nu\rho\sigma}~m_{\chi_j} [
 t_+^\mu p_{e^-}^\nu p_{e^+}^\rho s^{b,_\sigma} (t_-\cdot p_{\chi_i}) 
+t_-^\mu p_{e^-}^\nu p_{e^+}^\rho p_{\chi_i}^\sigma(t_+\cdot s^b)\nonumber\\
&& 
+ t_-^\mu t_+^\nu p_{e^+}^\rho s^{b,\sigma} (p_{e^-}\cdot p_{\chi_i})
+ t_-^\mu t_+^\nu p_{e^-}^\rho p_{\chi_i}^\sigma (p_{e^+}\cdot s^b)
]~,  
\label{eq:Sr2}
\end{eqnarray}
where $\varepsilon_{0123}=-1$.
The $e^{\pm}$ polarization vectors $t_\pm$ 
are given in Eq.~(\ref{eq:TPvec}), and the neutralino polarization
vectors are given in Eq.~(\ref{eq:polvec}), for details see Appendix~\ref{Momenta}. 

\medskip

In~\cite{Bartl:2005uh}, the properties of $\Sigma^b_{P,T}$~(\ref{eq:defsigbsum}) 
have been discussed in detail. In contrast to unpolarized and
longitudinally polarized beams,
it contains no contributions from pure selectron exchange,
$\Sigma^b_P(\tilde e_L \tilde e_L)_T$ and $\Sigma^b_P(\tilde e_R \tilde e_R)_T$.
However, transverse beam polarization leads to 
$\tilde e_L$-$\tilde e_R$ interference $\Sigma^b_P(\tilde e_L \tilde e_R)_T$, which
is absent for unpolarized and longitudinally 
polarized beams~\cite{Moortgat-Pick:1999di}. In addition,
there is no contribution $\Sigma^b_P(Z Z)_T$
owing to the Majorana character of the neutralinos.
Note that in the high energy limit $m_e/\sqrt{s}\to 0$,  
the neutralino polarization 
is proportional to the product of transverse beam polarizations
$\Sigma^b_{P,T}\propto{\cal P}_T^- {\cal P}_T^+$.

\medskip

The contributions to the amplitude squared, Eq.~(\ref{amplitude2}), from the 
neutralino decay are
\begin{eqnarray}
D&=&
 \frac{4g^4}{\cos^4\theta_W}(L_f^2+R_f^2)\Big\{
    2|O^{''L}_{nj}|^2( {p_{\chi_j}}\!\cdot\!{p_{\bar f}})[m_{\chi_j}^2-m_{\chi_i}^2
         +m_Z^2-2( {p_{\chi_j}}\!\cdot\!{p_{\bar f}})]
\nonumber\\ 
&&-|O^{''L}_{nj}|^2\frac{m_Z^2}{2}(m_{\chi_j}^2-m_{\chi_i}^2+m_Z^2)
    +[{\rm Re}(O^{''L}_{nj})^2 -{\rm Im}(O^{''L}_{nj})^2]m_{\chi_i}m_{\chi_j}m_Z^2
\Big\}~,
\label{Z:D}
\end{eqnarray}
and
\begin{eqnarray}
\Sigma_{D}^b&=&\frac{4g^4}{\cos^4\theta_W}(L_f^2-R_f^2)\Big\{
    |O^{''L}_{nj}|^2 m_{\chi_j}\Big[
(m_{\chi_j}^2-m_{\chi_i}^2-m_Z^2)({p_{\bar f}} \!\cdot\!{s_{\chi_j}^b})
-({p_{Z}} \!\cdot\!{s_{\chi_j}^b})( {p_{\chi_j}}\!\cdot\!{p_{\bar f}})
\nonumber\\ 
&&\!\!\!\!\!\!\!\!\!
+m_Z^2({p_{Z}} \!\cdot\!{s_{\chi_j}^b})
\Big]
-4{\rm Im}(O^{''L}_{nj}){\rm Re}(O^{''L}_{nj})m_{\chi_i}
\varepsilon_{\mu\nu\rho\sigma}s_{\chi_j}^{b,\mu} p_{Z}^\nu p_{\chi_j}^\rho p_{\bar f}^\sigma
\nonumber\\ 
&&\!\!\!\!\!\!\!\!\!
-[{\rm Re}(O^{''L}_{nj})^2 -{\rm Im}(O^{''L}_{nj})^2]m_{\chi_i}
\Big[(m_{\chi_j}^2-m_{\chi_i}^2+m_Z^2)({p_{\bar f}} \!\cdot\!{s_{\chi_j}^b})
-({p_{Z}} \!\cdot\!{s_{\chi_j}^b})( {p_{\chi_j}}\!\cdot\!{p_{\bar f}})
\Big]
\Big\}.\nonumber\\
\label{Z:SigmaD}
\end{eqnarray}
Different ways to calculate the squared amplitude $|T|^2$ for the decay chain
$\tilde\chi^0_j \to \tilde\chi^0_n Z^0$, followed by $Z^0 \to f \bar f$,
with complete spin correlations in the spin density matrix formalism,
have been given in detail in \cite{Bartl:2003ck,Bartl:2004ut,Kittel:2004rp}.
In contrast to previous calculations, we directly obtain the terms $D$ 
and $\Sigma_{D}^b$ for the entire
neutralino decay chain by summing 
over the internal polarization vectors $\epsilon^\mu_Z$
of the propagating $Z^0$ boson in the narrow width approximation.

\section{CP asymmetries 
%for transverse, longitudinal and/or unpolarized beams
        \label{asymmetry}}

For the combined process of neutralino production and 
decay~(\ref{production})--(\ref{decay_2B}),
the spin correlations  $\Sigma_P^b~ \Sigma_D^b$
in the amplitude squared $|T|^2$~(\ref{amplitude2}), allow us to 
define several CP sensitive observables. 
These are asymmetries in the 
azimuthal angular distribution of the final state fermions
from the $Z^0$ boson decay~(\ref{decay_2B}).
The CP asymmetries are based on different triple
product correlations, which
we classify in the following according to the beam polarization. 
We will define CP asymmetries for 
transversely as well as for longitudinally and/or unpolarized beams.

\subsection{Transverse beam polarizations
        \label{trans_asymmetry}}

%\subsection{CP asymmetries in  $e^+ e^- \to \tilde\chi^0_i  \tilde\chi^0_j\to
%  \tilde\chi^0_i \tilde\chi^0_n Z^0 \to  \tilde\chi^0_i \tilde\chi^0_n f \bar{f}$}

We define a CP observable for transverse beam polarizations 
which is based on the T-odd correlation
\begin{equation}
{\mathcal O}_T= \vec{t}_+\cdot (\hat{p}_f \times \hat{p}_{e^-}) 
(\vec{t}_- \cdot\hat{p}_f)
+ \vec{t}_- \cdot (\hat{p}_f \times \hat{p}_{e^-}) 
(\vec{t}_+ \cdot \hat{p}_f)=\sin(\eta-2 \phi_f)~,
\label{eq:Toddcor}
\end{equation}
where $\hat{p}_{e^-}$ is the unit vector of the $e^-$ beam, 
$\hat{p}_f$  the unit vector of the final fermion $f$ in 
$Z^0\to f \bar{f}$, $\phi_f$ is the
azimuthal angle of $f$, and the constant $\eta=\phi_++\phi_-$ is the
sum of the two azimuthal angles of the polarization vectors
of the positron beam, $\vec{t}_+$, and the electron beam, $\vec{t}_-$.
The equality ${\mathcal O}_T=\sin(\eta-2 \phi_f)$  
in Eq.~(\ref{eq:Toddcor}) follows from the specific parametrization 
of the momenta in the center-of-mass system,
for their definition see Appendix~\ref{Momenta}.
Note that the T-odd correlation ${\mathcal O}_T$ is contained in 
the kinematical factor $r^b_2$, Eq.~(\ref{eq:Sr2}), entering in the
neutralino polarization terms $\Sigma^b_{P,T}$,
given in Eqs.~(\ref{eq:SPZeL})--(\ref{eq:SPeLeR}). 
The CP asymmetry is then defined by
\begin{eqnarray} \label{asym}
{\mathcal A}^{\rm T}_{f}=
\frac{N[{\mathcal O}_T>0]-N[{\mathcal O}_T<0]}
{N[{\mathcal O}_T>0]+N[{\mathcal O}_T<0]}~,
\end{eqnarray}
where the number of events with ${\mathcal O}_T> 0 \,(< 0)$
is denoted by $N[{\mathcal O}_T> 0 \,(< 0)\,]$.
With the definition of the cross section, Eq.~(\ref{Z:crossection}),
${\mathcal L}~{\rm d}\sigma = {\rm d} N $, 
where ${\mathcal L}$ is the integrated luminosity,
and the amplitude squared~(\ref{amplitude2}) with the 
decompositions of the quantities $P$ and $\Sigma^b_P$, 
Eqs.~(\ref{eq:defpsum}) and (\ref{eq:defsigbsum}), 
we obtain for the CP asymmetry
\begin{eqnarray} \label{asym1}
{\mathcal A}^{\rm T}_{f}= 
\frac{\int {\rm sgn}[{\mathcal O}_T]~|T|^2~ d{\rm Lips}}
           {\int |T|^2 ~ d{\rm Lips}}=
\frac{\int {\rm sgn}[{\mathcal O}_T]~\Sigma_{P,T}^b~\Sigma_D^b~d{\rm Lips}^\prime}
{\int (P_{\rm unpol}+P_L)~D~d{\rm Lips}^\prime}~,
\end{eqnarray}
summed over $b=1,2,3$, and with
$d{\rm Lips}^\prime =d{\rm Lips}/(d s_{\chi_j}d s_{Z})$~(\ref{Lips}),
where we have already used the narrow width approximation for the
propagators~(\ref{narrowwidth}). 
In the numerator of ${\mathcal A}^{\rm T}_{f}$, Eq.~(\ref{asym1}), the
spin correlations $\Sigma_{P,T}^b \Sigma_D^b$ are only non-vanishing if
weighted with ${\rm sgn}[{\mathcal O}_T]$, as they include 
${\mathcal O}_T =\sin(\eta-2\phi_f)$. 
In the denominator of Eq.~(\ref{asym1}), we have used ${\int P_T~D~d{\rm Lips}^\prime}=0$ 
since the cross section is independent of the transverse beam polarizations.
Also the terms of the amplitude squared which depend on the 
$\tilde\chi^0_j$ polarization vanish for an integration over the entire phase space.  
The asymmetry ${\mathcal A}^{\rm T}_{f}$ is sensitive to CP violating couplings 
in the production only, which enter 
$\Sigma_{P,T}^b$, see Eq.~(\ref{eq:defsigbsum}).   
That means ${\mathcal A}^{\rm T}_{f}=0$ for the production of a pair of equal neutralinos
$e^+e^- \to \tilde\chi^0_i \tilde\chi^0_i$.
In addition, the asymmetry ${\mathcal A}^{\rm T}_{f}$
is proportional to the product of transverse beam polarizations
\begin{eqnarray} \label{asym_prop}
{\mathcal A}^{\rm T}_{f} \propto \Sigma^b_{P,T} \propto {\cal P}_T^- {\cal P}_T^+~,
\end{eqnarray}
and thus vanishes if one beam is unpolarized.

\medskip

For the measurement of ${\mathcal A}^{\rm T}_{f}=-{\mathcal A}^{\rm T}_{\bar f}$, 
the charges and the flavors of the final fermions $f$ and $\bar f$
have to be  distinguished. For $f=e,\mu$ this will be 
possible on an event by event basis. 
For $f=\tau$ it will be possible after taking into account
corrections due to the reconstruction of the $\tau$ 
momentum. For $f=q$ the distinction of the quark 
flavors should be possible by flavor tagging of the heavy 
quarks $q=b,c$~\cite{XellaHansen:2003sw,Abe:2001dr}. 
The distinction of the heavy quark charges can be accomplished
with very good precision in the case of semi-leptonic decays 
of the hadrons. For the majority of $b$ and $c$-jets, it also 
can be accomplished by the reconstruction of the vertex charge in 
the cases where the hadrons decay 
non-leptonically~\cite{Abe:2001dr,TDR,private}.
However, $b$ and $c$ tagging will be essential.

\medskip

The asymmetry for final quarks ${\mathcal A}^{\rm T}_{q}$ is always larger than
that for final leptons ${\mathcal A}^{\rm T}_{\ell}$, due to the dependence of ${\mathcal A}^{\rm T}_{f}$ 
on the $Z^0 f \bar f$ couplings~\cite{Bartl:2003ck,Bartl:2004ut,Kittel:2004rp}
\begin{eqnarray} \label{propto}
{\mathcal A}^{\rm T}_{f}\propto \frac{R_f^2-L_f^2}{R_f^2+L_f^2}
\quad\Rightarrow\quad
{\mathcal A}^{\rm T}_{b(c)}&=&
\frac{R_{\ell}^2+L_{\ell}^2}
     {R_{\ell}^2-L_{\ell}^2}
\frac{R_{b(c)}^2-L_{b(c)}^2}
     {R_{b(c)}^2+L_{b(c)}^2}~
          {\mathcal A}^{\rm T}_{\ell}\\[3mm]
\quad\Rightarrow\quad
{\mathcal A}^{\rm T}_{b(c)} &\simeq& 6.3~(4.5)\times{\mathcal A}^{\rm T}_{\ell}~,
\label{propto1}
\end{eqnarray}
which follows from Eqs.~(\ref{Z:D}), (\ref{Z:SigmaD}) and~(\ref{asym}).

\subsection{Longitudinal or unpolarized beams
        \label{long_asymmetry}}

For longitudinal or unpolarized beams,
the triple product~\cite{Kizukuri:1990iy}
\begin{eqnarray} \label{eq:Toddcor2}
{\mathcal O}_L=\vec{p}_{e^-}\cdot (\vec{p}_f \times \vec{p}_{\bar f})~,
\end{eqnarray}
can be used to define the CP asymmetry 
\begin{eqnarray} \label{asym2}
{\mathcal A}^{\rm L}_{f}= 
\frac{\int {\rm sgn}[{\mathcal O}_L]~|T|^2~ d{\rm Lips}}
           {\int |T|^2 ~ d{\rm Lips}}=
\frac{\int {\rm sgn}[{\mathcal O}_L]~(\Sigma_{P,{\rm unpol}}^b
+\Sigma_{P,L}^b)~\Sigma_D^b~d{\rm Lips}^\prime}
{\int (P_{\rm unpol}+P_L)~D~d{\rm Lips}^\prime}~.
\end{eqnarray}
This CP asymmetry has been analyzed in detail 
in~\cite{Bartl:2004ut}. Due to the different angular dependence
of the triple product ${\mathcal O}_L$, Eq.~(\ref{eq:Toddcor2}), and the 
CP-odd terms in $\Sigma_{P,T}^b$, we have 
${\int {\rm sgn}[{\mathcal O}_L]~\Sigma_{P,T}^b~\Sigma_D^b~d{\rm
    Lips}^\prime}=0$.
This means that the asymmetry ${\mathcal A}^{\rm L}_{f}$ does not depend on
the degree of transverse beam polarization.
In contrast to the asymmetry ${\mathcal A}^{\rm T}_{f}$, Eq.~(\ref{asym1}), the 
asymmetry ${\mathcal A}^{\rm L}_{f}$ is also sensitive to the CP violating couplings
in the neutralino decay
$\tilde{\chi}^0_j \to \tilde{\chi}^0_n Z^0$~\cite{Bartl:2004ut}.
Due to the dependence of ${\mathcal A}^{\rm L}_{f}$ 
on the $Z^0 f \bar f$ couplings,
we again have~\cite{Bartl:2003ck,Bartl:2004ut,Kittel:2004rp}
\begin{eqnarray} \label{propto2}
{\mathcal A}^{\rm L}_{b(c)} &\simeq& 6.3~(4.5)\times{\mathcal A}^{\rm L}_{\ell}~.
\end{eqnarray}

\subsection{Measurability of the CP asymmetries
        \label{significance}}

For a measurement of non-vanishing phases in the neutralino sector 
at the ILC, not only large CP asymmetries 
but also large cross sections are required.
Since the cross section appears in the denominator of
the CP asymmetries 
${\mathcal A}^{\rm T}_{f}$, Eq.~(\ref{asym1}), and 
${\mathcal A}^{\rm L}_{f}$, Eq.~(\ref{asym2}),
we are often confronted with a situation where large cross sections 
lead to small asymmetries and vice versa.
In order to estimate whether the CP asymmetries 
can be measured at the ILC, 
we consider their statistical
significances~\cite{Bartl:2003ck,Bartl:2004ut,Kittel:2004rp}.

\medskip

The significance for the 
CP asymmetry ${\mathcal A}^{\rm T}_{f}$ with transversely polarized beams
is defined by
\begin{equation}
\label{estimate}
S^{\rm T}_f= |{\mathcal A}^{\rm T}_{f}|\sqrt{{\mathcal L}~
\sigma(e^+e^- \to \tilde\chi^0_i \tilde\chi^0_j)_{\rm unpol} \,
{\rm BR}(\tilde\chi^0_j \to \tilde\chi^0_n Z^0) \, 
{\rm BR}(Z^0 \to f\bar{f})}~.
\end{equation} 
Note that only the unpolarized cross section 
$\sigma(e^+e^- \to \tilde\chi^0_i \tilde\chi^0_j)_{\rm unpol}$
appears in Eq.~(\ref{estimate}), since the cross section is independent of
transversal degrees of beam polarizations.
We define the statistical significances for the asymmetries 
${\mathcal A}^{\rm L}_{f}$ by
\begin{equation}
\label{estimate1}
S^{\rm L}_f= |{\mathcal A}^{\rm L}_{f}|\sqrt{{\mathcal L}~
\sigma(e^+e^- \to \tilde\chi^0_i \tilde\chi^0_j) \,
{\rm BR}(\tilde\chi^0_j \to \tilde\chi^0_n Z^0) \, 
{\rm BR}(Z^0 \to f\bar{f})}~.
\end{equation}
Here, 
the cross section depends on the
degree of longitudinal electron and positron beam polarizations.
Note that $S^{\rm T}_f$ and $S^{{\rm L}}_f$, in short $S_{f}$,
are larger for $f=b,c$ than for $f=\ell=e,\mu,\tau$ 
with~\cite{Bartl:2003ck,Bartl:2004ut,Kittel:2004rp}
\begin{equation}\label{Sigmarel}
S_{b} \simeq 7.7\times S_{\ell} \quad  
{\rm and} \quad S_{c} \simeq 4.9\times S_{\ell}~,
\end{equation}
caused by larger asymmetries~(\ref{propto}), (\ref{estimate}), and by
larger branching ratios
${\rm BR}(Z^0\to b\bar b) \simeq1.5\times\sum_\ell{\rm BR}(Z^0\to\ell\bar\ell)$,
and
${\rm BR}(Z^0\to c\bar c) \simeq1.2\times\sum_\ell{\rm BR}(Z^0\to\ell\bar\ell)$,
$\ell = e, \mu,\tau$, with 
$\sum_\ell{\rm BR}(Z^0\to\ell\bar\ell)\simeq0.1$ \cite{Yao:2006px}.

\medskip

For an ideal detector, a significance of, e.g., $S_{f} = 1$ implies
that ${\mathcal A}_{f}$ 
can be measured at the statistical 68\% confidence level.
However, the definition of our theoretical significance 
is based on statistics only,
and does not include detector and particle reconstruction
efficiencies.
Our significances are thus 
upper bounds to judge on the feasibility to measure
the CP asymmetries.
In order to predict the absolute values of confidence levels,
detailed Monte Carlo analyses
including detector and background 
simulations with particle identification and reconstruction
efficiencies would be required.
A Monte Carlo analysis for a CP asymmetry in the 
production and
decay of neutralinos with longitudinal polarized beams
has been carried out in~\cite{Aguilar-Saavedra:2004dz}.
However, such an analysis is beyond the scope
of the present work.
We only estimate how the
detection rates for $b$- and $c$-quark jets have to be modified. 
Using vertex detectors, flavor 
tagging of $b$- and $c$-quarks is possible
and the corresponding 
efficiencies and purities have been studied~\cite{TDR}. 
It has been shown~\cite{XellaHansen:2003sw},
that $b$-quarks ($c$-quarks) can be identified with an efficiency
of $50\%$ ($50\%$) at a purity of $90$\% ($80\%$) for $e^+e^- \to q\bar{q}$ 
at $\sqrt{s}=500$~GeV. This would lead to a reduction of our statistical 
significance $S_b$ ($S_c$) by a factor of $0.64$ ($0.57$).

\begin{figure}[p]
\setlength{\unitlength}{1mm}
\begin{center}
\begin{picture}(150,200)
 \put(-53,-70){\mbox{\epsfig{figure=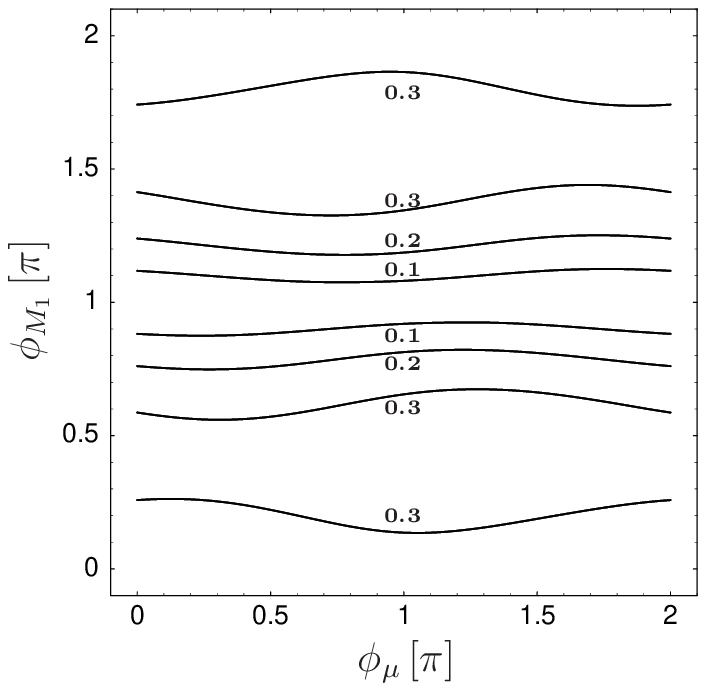,
height=31cm,width=19.4cm}}}
 \put(27,-70){\mbox{\epsfig{figure=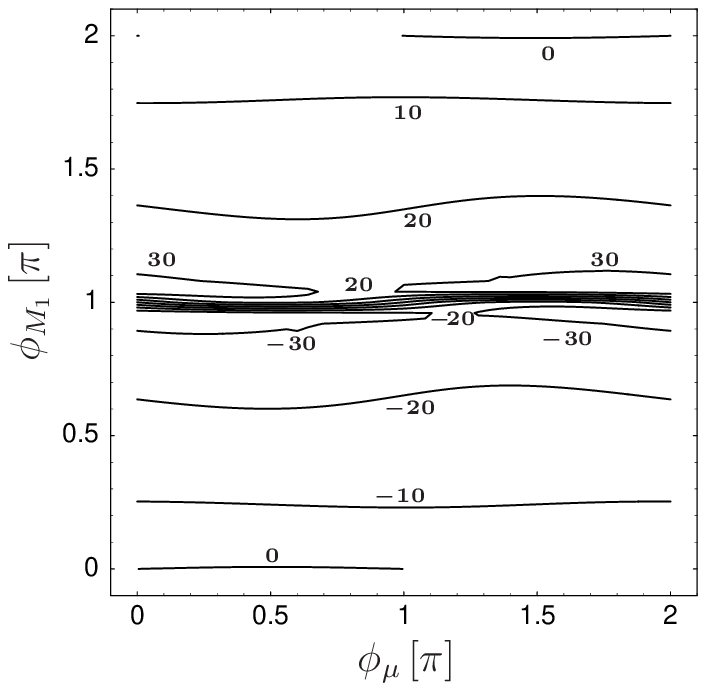,
height=31cm,width=19.4cm}}}
 \put(-53,-153){\mbox{\epsfig{figure=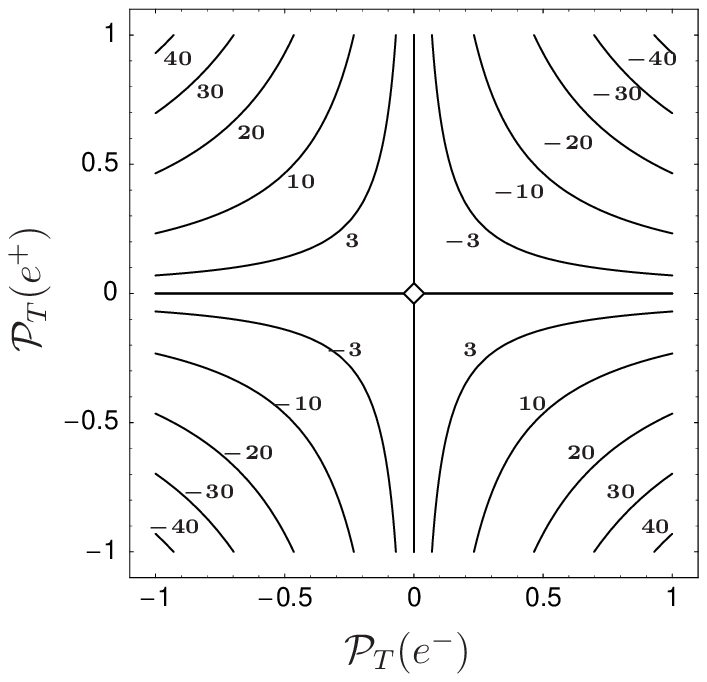,
height=31cm,width=19.4cm}}}
 \put(27,-153){\mbox{\epsfig{figure=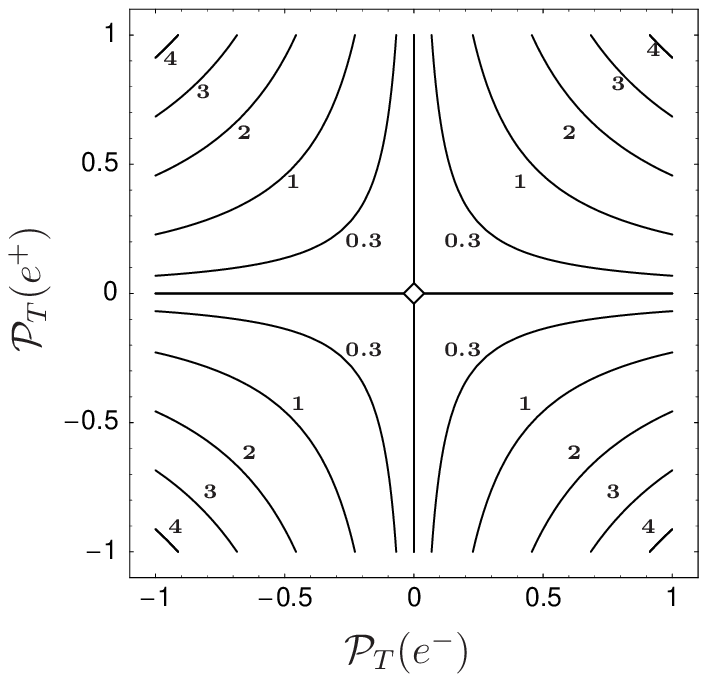,
height=31cm,width=19.4cm}}}
\put(8,207){\mbox{(a) $ \quad
\sigma(e^+e^- \to \tilde{\chi}^0_1 \tilde{\chi}^0_1 b \bar{b}) \,[{\rm fb}] $}}
\put(88,207){\mbox{(b) $ \qquad \qquad {\mathcal A}^{\rm T}_b \, [\%] $}}
\put(8,123){\mbox{(c) $ \qquad \qquad {\mathcal A}^{\rm T}_b \, [\%] $}}
\put(88,123){\mbox{(d) $ \qquad S^{\rm T}_b= 
|{\mathcal A}^{\rm T}_b|\sqrt{\sigma \cdot {\mathcal L}}$}}
\end{picture}
\end{center}
\vspace{-5cm}
\caption{
    Neutralino production
    $e^+e^- \to \tilde{\chi}^0_1 \tilde{\chi}^0_3$
    and decay
    $\tilde{\chi}^0_3 \to \tilde{\chi}^0_1 Z^0$, $Z^0\to b\bar{b}$,
    at $\sqrt s=500$~GeV for Scenario A given in Table \ref{tab1}.
    Phase dependence of 
    (a)~the cross section $\sigma(e^+e^- \to \tilde{\chi}^0_1 \tilde{\chi}^0_1 b
    \bar{b})\equiv\sigma(e^+e^- \to \tilde{\chi}^0_1 \tilde{\chi}^0_3)\cdot
    {\rm BR}(\tilde{\chi}^0_3 \to \tilde{\chi}^0_1 Z^0)\cdot
    {\rm BR}(Z^0 \to \bar{b} b)$, and
    (b)~the CP asymmetry ${\mathcal A}^{\rm T}_b$ for
    transverse beam polarizations
    ${\mathcal P}_T(e^-)=0.9$, ${\mathcal P}_T(e^+)=0.6$.
    Contour lines in the ${\mathcal P}_T(e^-)$--${\mathcal P}_T(e^+)$ plane for
    (c)~the CP asymmetry ${\mathcal A}^{\rm T}_b$, and
    (d)~the statistical significance $S^{\rm T}_b$,
    with $\phi_{M_1}=\frac{3\pi}{4}$, $\phi_\mu=0$, and
    an integrated luminosity ${\mathcal L}=500$~fb$^{-1}$.
}
\label{N1N3_I}
\end{figure}

\section{Numerical results
        \label{Numerical results}}

We present numerical results for the CP asymmetry ${\mathcal A}^{\rm T}_f$, Eq.~(\ref{asym}),
with transverse beam polarizations,
and for the CP asymmetry ${\mathcal A}^{\rm L}_f$, Eq.~(\ref{asym2}),
with longitudinal beam polarizations.
For neutralino production
$e^+e^- \to \tilde{\chi}^0_1 \tilde{\chi}^0_{2(3)}$ and decay 
$\tilde{\chi}^0_{2(3)} \to \tilde{\chi}^0_1 Z^0$, $Z^0\to b \bar{b}$,
we study the dependences of ${\mathcal A}^{\rm T}_b$ and the cross section
on the gaugino mass parameter $M_1=|M_1|e^{i\phi_{M_1}}$ and
on the higgsino mass parameter $\mu=|\mu|e^{i\phi_{\mu}}$,
as well as on the beam polarizations 
${\mathcal P}(e^-)$ and ${\mathcal P}(e^+)$.
Throughout this study we take $\sqrt{s}=500$~GeV.
The values of the CP asymmetries for final leptons ${\mathcal A}^{\rm T}_\ell$,
and those for final $c$-quarks 
${\mathcal A}^{\rm T}_c$, are related to ${\mathcal A}^{\rm T}_b$ by 
Eq.~(\ref{propto}).
Finally we compute cross sections and asymmetries which are
accessible with longitudinally polarized beams, and compare them 
with the results for transversely polarized beams.

\medskip

For the calculation of the neutralino widths and 
branching ratios,
we include the two-body decays~\cite{Kittel:2004rp}
\begin{eqnarray}
        \tilde\chi^0_i &\to& \tilde \ell_{n} \ell,~ 
        \tilde\nu_{\ell} \nu_{\ell},~
        \tilde\chi^0_1 Z^0,~
        \tilde\chi^{\mp}_1 W^{\pm},~
        \tilde\chi^0_1 H_1^0~,
\end{eqnarray}
with $n=R,L$  for $\ell=e,\mu$, and $n=1,2$ for $\ell=\tau$.
The Higgs mass parameter is chosen $m_{A}=1$~TeV.  
With such a high value for $m_{A}$
explicit CP violation in the Higgs sector is not important
for the lightest Higgs state $H_1^0$ \cite{Carena:2000yi}.
In the stau sector, we fix the trilinear scalar coupling parameter 
$A_{\tau}=250$ GeV. To reduce the number of free parameters, we take
$|M_1|=(5/3)~M_2 \tan^2\theta_W$,
inspired by gaugino mass unification.

\begin{table}[H]
\begin{center}
\begin{tabular}{|c||c|c|c|c|c|c||c|c|c|c|} \hline
 Scenario 
& \multicolumn{1}{c|}{$M_2$} 
& \multicolumn{1}{c|}{$|\mu|$}  
& \multicolumn{1}{c|}{$\tan{\beta}$}  
& \multicolumn{1}{c|}{$m_{\tilde{e}_L}$}
& \multicolumn{1}{c||}{$m_{\tilde{e}_R}$}
& \multicolumn{1}{c|}{$m_{\chi_1^0}$}
& \multicolumn{1}{c|}{$m_{\chi_2^0}$}
& \multicolumn{1}{c|}{$m_{\chi_3^0}$}
& \multicolumn{1}{c|}{$m_{\chi_1^\pm}$}
\\\hline\hline
 A 
& \multicolumn{1}{c}{380} 
& \multicolumn{1}{c}{120}  
& \multicolumn{1}{c}{50} 
& \multicolumn{1}{c}{250}
& \multicolumn{1}{c||}{210}
& \multicolumn{1}{c}{107}
& \multicolumn{1}{c}{119}
& \multicolumn{1}{c}{205}
& \multicolumn{1}{c|}{114}
\\\hline
 B 
& \multicolumn{1}{c}{250} 
& \multicolumn{1}{c}{200}  
& \multicolumn{1}{c}{7} 
& \multicolumn{1}{c}{240}
& \multicolumn{1}{c||}{220}
& \multicolumn{1}{c}{119}
& \multicolumn{1}{c}{163}
& \multicolumn{1}{c}{213}
& \multicolumn{1}{c|}{161}
\\\hline
 C 
& \multicolumn{1}{c}{280} 
& \multicolumn{1}{c}{340}  
& \multicolumn{1}{c}{3} 
& \multicolumn{1}{c}{350}
& \multicolumn{1}{c||}{250}
& \multicolumn{1}{c}{141}
& \multicolumn{1}{c}{236}
& \multicolumn{1}{c}{349}
& \multicolumn{1}{c|}{234}
\\\hline
\end{tabular}\\[0.5ex]
\caption{\label{tab1}
Input parameters $M_2$, $|\mu|$, $\tan\beta$, 
$m_{\tilde{e}_L}$, and $m_{\tilde{e}_R}$
for Scenarios A, B, and C, with
$|M_1|=(5/3)~M_2 \tan^2\theta_W$.
The neutralino and chargino masses are calculated with
$\phi_{M_1}=\frac{3\pi}{4}$ and $\phi_{\mu}=0$.
All mass parameters are given in GeV.}
\end{center}
\end{table}

\subsection{Neutralino $\tilde\chi^0_1 \tilde\chi^0_3$ production and decay
$\tilde\chi^0_3 \to \tilde\chi^0_1 Z^0 \to  \tilde\chi^0_1 b\bar{b}$}

In Fig.~\ref{N1N3_I}a, we show contour lines of the
cross section $\sigma(e^+e^- \to \tilde{\chi}^0_1 \tilde{\chi}^0_1 b
\bar{b})\equiv \sigma(e^+e^- \to \tilde{\chi}^0_1 \tilde{\chi}^0_3)\cdot
{\rm BR}(\tilde{\chi}^0_3 \to \tilde{\chi}^0_1 Z^0)\cdot
{\rm BR}(Z^0 \to \bar{b} b)$ in the $\phi_\mu$--$\phi_{M_1}$ plane 
for Scenario A, see Table~\ref{tab1}. 
The production cross section $\sigma(e^+e^- \to \tilde{\chi}^0_1
\tilde{\chi}^0_3)_{\rm unpol}$ as well as the decay branching ratio 
${\rm BR}(\tilde{\chi}^0_3 \to \tilde{\chi}^0_1 Z^0)$ 
(which varies between 5\% and 15\%) strongly depend on 
$\phi_{M_1}$ but only mildly on $\phi_\mu$, and their minimum values are 
obtained for $\phi_{M_1}=\pi$ and $\phi_\mu=0$. 
In Fig.~\ref{N1N3_I}b we show the phase dependence of the 
CP asymmetry ${\mathcal A}^{\rm T}_b$, Eq.~(\ref{asym}). 
The asymmetry ${\mathcal A}^{\rm T}_b$ reaches its maximum of about $36\%$ 
at $\phi_{M_1}=1.05\pi$ and $\phi_{\mu}=1.8\pi$. 
The maximum of the asymmetry ${\mathcal A}^{\rm T}_b$
is close to $\phi_{M_1}=\pi$. The reason for this is that the cross section, 
which is the denominator of the asymmetry, has a minimum there.      
We remark that the CP asymmetry is asymmetric with respect to
the transformations $\phi_{M_1}\to 2\pi-\phi_{M_1}$ and $\phi_\mu\to
2\pi-\phi_\mu$, while any CP-even observable, e.g. the 
cross section in Fig.~\ref{N1N3_I}a, must be symmetric under
these transformations. 
Fig.~\ref{N1N3_I}c shows the dependence of  
${\mathcal A}^{\rm T}_b \propto{\mathcal P}_T(e^-) \cdot {\mathcal P}_T(e^+)$
on the degrees of transverse 
polarizations of the electron and positron beams for Scenario A with
$\phi_{M_1}=\frac{3\pi}{4}$ and $\phi_{\mu}=0$.
The maximum value of the asymmetry 
${\mathcal A}^{\rm T}_b\approx \pm 40\%$
is attained for the maximal degree of polarizations ${\mathcal P}_T(e^+)=\pm 1$ and
${\mathcal P}_T(e^-)=\mp 1$. 
In Fig.~\ref{N1N3_I}d we show
the statistical significance $S_b$, Eq.~(\ref{estimate}), 
for an integrated luminosity of ${\mathcal L}=500$~fb$^{-1}$.
A measurement of the asymmetry thus requires a high degree of 
the beam polarizations. In the following 
we assume that a polarization 
of 90\% for the electron beam, and 60\% for the positron beam
is feasible, and study the $|\mu|$--$M_2$ dependence of
the asymmetry ${\mathcal A}^{\rm T}_b$.

\medskip

\begin{figure}[p]
\setlength{\unitlength}{1mm}
\begin{center}
\begin{picture}(150,200)
 \put(-53,-70){\mbox{\epsfig{figure=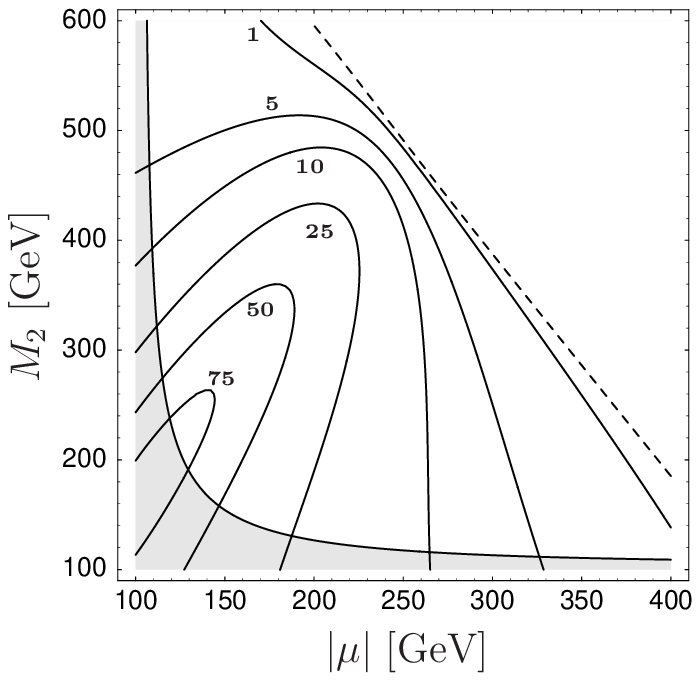,
height=31cm,width=19.4cm}}}
 \put(27,-70){\mbox{\epsfig{figure=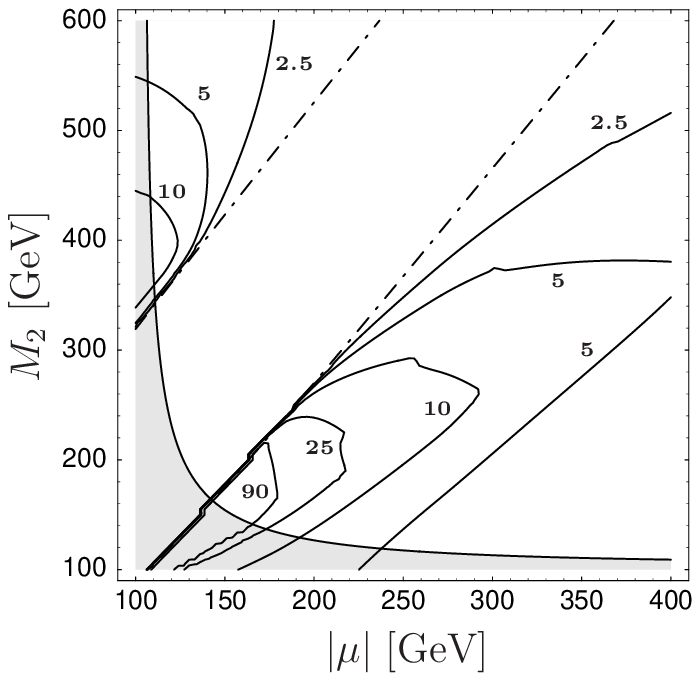,
height=31cm,width=19.4cm}}}
 \put(-53,-155){\mbox{\epsfig{figure=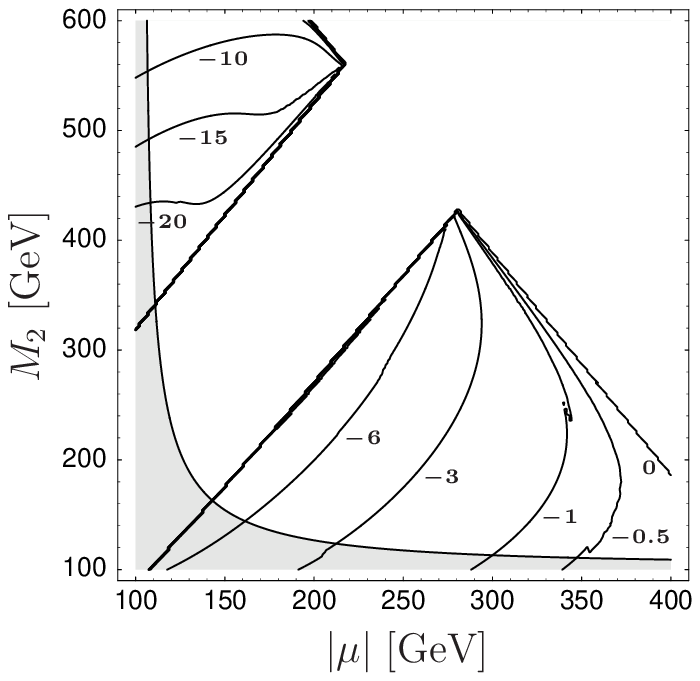,
height=31cm,width=19.4cm}}}
 \put(27,-155){\mbox{\epsfig{figure=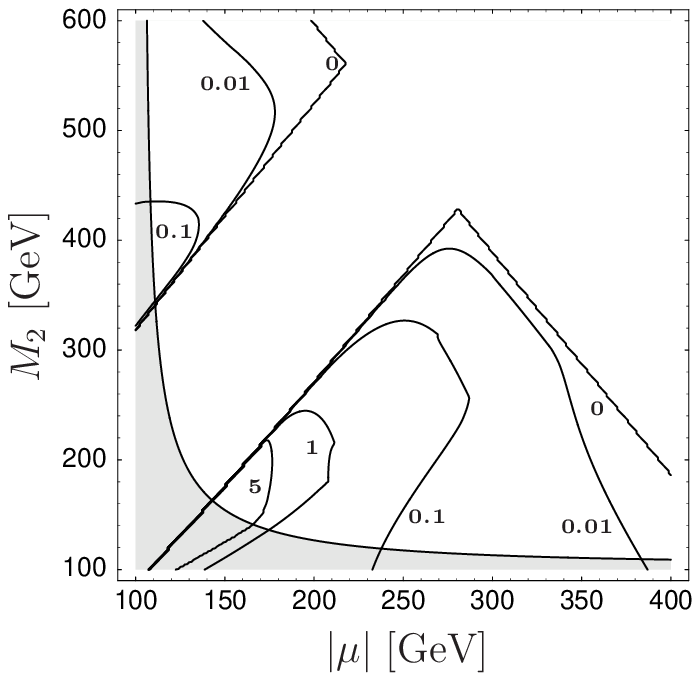,
height=31cm,width=19.4cm}}}
\put(8,205){\mbox{(a) $ \quad
\sigma(e^+e^- \to \tilde{\chi}^0_1 \tilde{\chi}^0_3) \,[{\rm fb}] $}}
\put(88,205){\mbox{(b) $ \qquad {\rm BR}(\tilde{\chi}^0_3 \to \tilde{\chi}^0_1 Z^0)$}}
\put(8,120){\mbox{(c) $\qquad \qquad {\mathcal A}^{\rm T}_b \, [\%] $}}
\put(88,120){\mbox{(d) $ \quad \sigma(e^+e^- \to \tilde{\chi}^0_1
\tilde{\chi}^0_1 b \bar{b}) \,[{\rm fb}] $}}
\end{picture}
\end{center}
\vspace{-5cm}
\caption{
    Contour lines in the $|\mu|$--$M_2$ plane of  
    (a)~the production cross section
    $\sigma(e^+e^- \to \tilde\chi^0_1 \tilde\chi^0_3)$,  
    (b)~the branching ratio ${\rm BR}(\tilde{\chi}^0_3 \to \tilde{\chi}^0_1 Z)$,
    (c)~the CP asymmetry ${\mathcal A}^{\rm T}_b$,
    and (d)~the total cross section 
    $\sigma(e^+e^- \to \tilde{\chi}^0_1 \tilde{\chi}^0_1 b \bar{b})$,
    for Scenario A given in Table \ref{tab1},
    with $\phi_{M_1}=\frac{3 \pi}{4}$, $\phi_\mu=0$, at
    $\sqrt s=500$~GeV with
    transverse beam polarizations ${\mathcal P}_T(e^-)=0.9$, ${\mathcal P}_T(e^+)=0.6$.
    In the gray shaded area $m_{\chi^\pm_1}<104$~GeV. 
    The dashed line in~(a) indicates the kinematical limit 
    $m_{\chi_1^0} +m_{\chi_3^0}=\sqrt s $, and the dashed-dotted line
    in~(b) indicates the limit $m_{\chi_3^0}=m_{\chi_1^0}+m_{Z^0}$.
}
\label{N1N3_muM2}
\end{figure}
 
In Fig.~\ref{N1N3_muM2}a, we show contour lines of the 
production cross section $\sigma(e^+e^- \to \tilde{\chi}^0_1
\tilde{\chi}^0_3)$ in the $|\mu|$--$M_2$ plane,
with $\phi_{M_1}=\frac{3\pi}{4}$ and $\phi_\mu=0$.
The other MSSM parameters are fixed
as defined in Scenario A of Table~\ref{tab1}.
In Fig.~\ref{N1N3_muM2}b we show the contour lines of the branching
ratio ${\rm BR}(\tilde{\chi}^0_3 \to \tilde{\chi}^0_1 Z^0)$. It is as large
as $90\%$ for $M_2 \sim |\mu|\sim 170$~GeV. 
For $|\mu|\gsim M_2$ and $|\mu|\gsim 200$~GeV the decay 
$\tilde{\chi}^0_3 \to \tilde{\chi}_1^\pm W^\mp$ dominates. 
Fig.~\ref{N1N3_muM2}c shows the corresponding CP asymmetry 
${\mathcal A}^{\rm T}_b$ in the $|\mu|$--$M_2$ plane.
The asymmetry  ${\mathcal A}^{\rm T}_b$ attains large values 
in the parameter region where the lighter neutralino
states $\tilde\chi^0_{1,2}$ are higgsino-like. 
For instance, for $|\mu| \approx 140$~GeV and
$M_2 \approx 420$~GeV, the asymmetry is ${\mathcal A}^{\rm T}_b\approx-20$\%. 
In addition, we show in Fig.~\ref{N1N3_muM2}d the cross section
$\sigma(e^+e^- \to \tilde{\chi}^0_1 \tilde{\chi}^0_1 b \bar{b})$
$(\equiv \sigma(e^+e^- \to \tilde{\chi}^0_1 \tilde{\chi}^0_3)\cdot
{\rm BR}(\tilde{\chi}^0_3 \to \tilde{\chi}^0_1 Z^0)\cdot
{\rm BR}(Z^0 \to \bar{b} b))$. For $|\mu| \approx 150$~GeV and
$M_2 \approx 170$~GeV,  we obtain the largest values 
up to $7$~fb, where the asymmetry
${\mathcal A}^{\rm T}_b$ reaches $8.5\%$. 
With these values of ${\mathcal A}^{\rm T}_b$ and $\sigma$,
the statistical significance, Eq.~(\ref{estimate}), is $S^{\rm T}_b \approx 5$, for
an integrated luminosity of ${\mathcal L}=500$~fb$^{-1}$.

\medskip

Now we discuss ${\mathcal A}^{\rm T}_b$ in
neutralino production $e^+e^- \to \tilde\chi^0_1 \tilde\chi^0_3$
and decay $\tilde\chi^0_3 \to \tilde\chi^0_1 Z^0$,
$Z^0 \to \bar{b} b$, 
for Scenario B, see Table \ref{tab1},
with a small value for $\tan\beta=7$.
In this scenario, the neutralino $\tilde\chi^0_3$ is a strong mixture of gaugino and 
higgsino components, and 
${\rm BR}(\tilde\chi^0_3\to \tilde\chi^0_1 Z^0)=100\%$.
In Fig.~\ref{N1N3_II}a we show contour lines of the production cross section
$\sigma(e^+e^- \to \tilde{\chi}^0_1 \tilde{\chi}^0_3)$ in the
$\phi_\mu$--$\phi_{M_1}$ plane.
The production cross section ranges from $25$~fb, 
for $\phi_{M_1}=\pi$ and $\phi_{\mu}=0$, to $54$~fb, 
for $\phi_{M_1}=\phi_{\mu}=0$. 
In Fig.~\ref{N1N3_II}b, we show the total 
cross section $\sigma(e^+e^- \to \tilde{\chi}^0_1 \tilde{\chi}^0_1 b \bar{b})$,
which varies between $4$~fb and $8$~fb. 
Contour lines for the CP asymmetry ${\mathcal A}^{\rm T}_b$ are shown in Fig.~\ref{N1N3_II}c. 
As can be seen, ${\mathcal A}^{\rm T}_b$ can attain large values even for CP phases
close to the CP conserving case.
For instance, for $\phi_{\mu}=0$ and $\phi_{M_1}=0.8\pi$,  
the asymmetry reaches about $-10\%$.  
The corresponding statistical significance,
shown in Fig.~\ref{N1N3_II}d, reaches $S^{\rm T}_b=5$, 
for ${\mathcal L}=500$~fb$^{-1}$.

\medskip

\subsubsection{Comparison of transversely and longitudinally polarized beams}

\begin{figure}[t]
\setlength{\unitlength}{1mm}
\begin{center}
\begin{picture}(150,200)
 \put(-53,-70){\mbox{\epsfig{figure=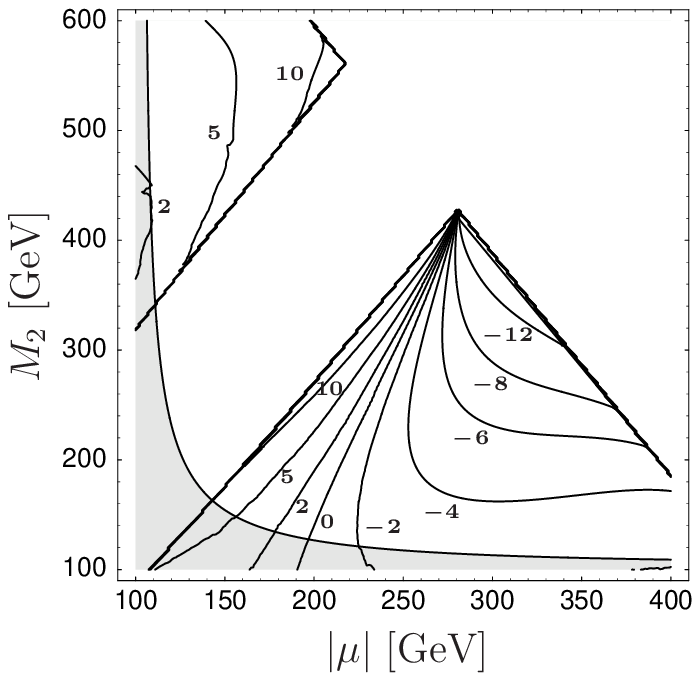,
height=31cm,width=19.4cm}}}
\put(27,-70){\mbox{\epsfig{figure=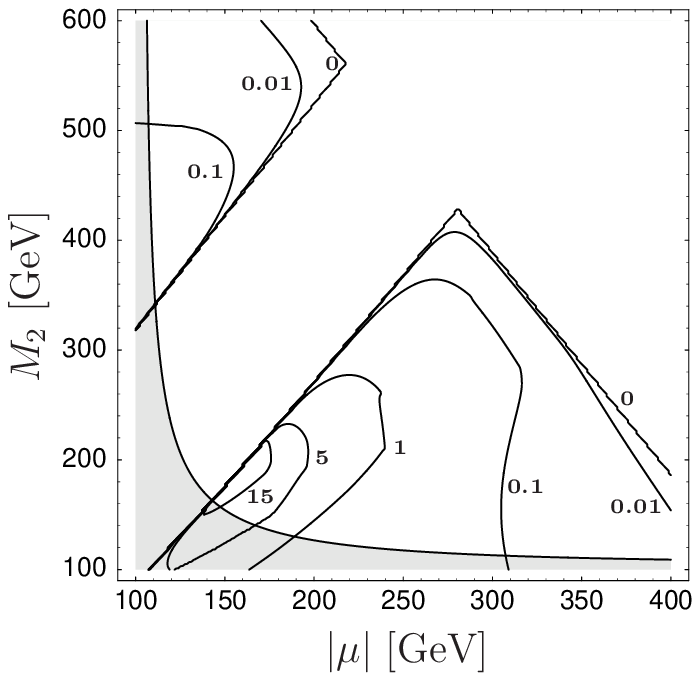,
height=31cm,width=19.4cm}}}
\put(8,205){\mbox{(a) $ \qquad \qquad {\mathcal A}^{\rm L}_b \, [\%] $}}
\put(88,205){\mbox{(b) $ \quad
\sigma(e^+e^- \to \tilde{\chi}^0_1 \tilde{\chi}^0_1 b \bar{b}) \,[{\rm fb}] $}}
\end{picture}
\end{center}
\vspace{-13.5cm}
\caption{
    Neutralino production
    $e^+e^- \to \tilde{\chi}^0_1 \tilde{\chi}^0_3$
    and decay
    $\tilde{\chi}^0_3 \to \tilde{\chi}^0_1 Z^0$, $Z^0\to b\bar{b}$
    at $\sqrt s=500$~GeV with longitudinal beam polarizations
    ${\mathcal P}_L(e^-)=0.9$ and ${\mathcal P}_L(e^+)=-0.6$.
    Contour lines in the $|\mu|$--$M_2$ plane for
    (a)~the CP asymmetry ${\mathcal A}^{\rm L}_b$, and 
    (b)~the total cross section 
    $\sigma(e^+e^- \to \tilde{\chi}^0_1 \tilde{\chi}^0_1 b \bar{b})$,
    for Scenario~A, given in Table~\ref{tab1}.
    The branching ratio
    ${\rm BR}(\tilde{\chi}^0_3 \to \tilde{\chi}^0_1 Z^0)$
    is shown in Fig.~\ref{N1N3_muM2}(b).
}
\label{N1N3_muM2_LongPol}
\end{figure}

In this subsection we compare the asymmetry 
${\mathcal A}^{\rm T}_b$, Eq.~(\ref{asym}), 
that is obtained with transverse beam polarization 
with the asymmetry ${\mathcal A}^{\rm L}_b$, Eq.~(\ref{asym2}), 
for longitudinally or unpolarized beams.
Fig.~\ref{N1N3_muM2_LongPol} shows the contours of the asymmetry
${\mathcal A}^{\rm L}_b$ and the total cross section 
$\sigma(e^+e^- \to \tilde\chi^0_1 \tilde\chi^0_1 b \bar{b})$
for longitudinally polarized beams in the $|\mu|$--$M_2$ plane
for scenario A, see Table \ref{tab1}.
The asymmetry ${\mathcal A}^{\rm L}_b$ can reach
$\pm 10\%$, while the total cross section can reach $15$~fb.
By comparing these results for longitudinal beam polarizations with
the results of Fig.~\ref{N1N3_muM2} for transverse beam polarizations,
we find that in general
for higgsino-like scenarios the statistical significance 
of asymmetry ${\mathcal A}^{\rm T}_b$ 
is larger than
that for asymmetry ${\mathcal A}^{\rm L}_b$.
The reason is that for higgsino-like scenarios the leading contribution to the
asymmetry ${\mathcal A}^{\rm T}_b$ is due to the $\tilde{e}_L$--$\,\tilde{e}_R$ 
interference term, Eq.~(\ref{eq:SPeLeR}), which is absent  
for longitudinally polarized beams.
Clearly, for larger values of the selectron masses
$m_{\tilde{e}_L}$ and $m_{\tilde{e}_R}$, the significance of
the asymmetry ${\mathcal A}^{\rm T}_b$ is reduced due to a suppression
of the $\tilde{e}_L$--$\,\tilde{e}_R$ interference term, 
as these masses enter in the propagators, see 
Fig.~\ref{Fig:FeynProd}.
On the other hand, for gaugino-like scenarios, the 
statistical significance of asymmetry ${\mathcal A}^{\rm L}_b$,
is larger than that for
${\mathcal A}^{\rm T}_b$, since then the 
$\tilde{e}_R$--$\,\tilde{e}_R$ and the
$\tilde{e}_L$--$\,\tilde{e}_L$ terms, which are only present
for unpolarized and longitudinally polarized beams, give
the dominant contribution to ${\mathcal A}^{\rm L}_b$.
   
\medskip

\begin{figure}[p]
\setlength{\unitlength}{1mm}
\begin{center}
\begin{picture}(150,200)
 \put(-53,-70){\mbox{\epsfig{figure=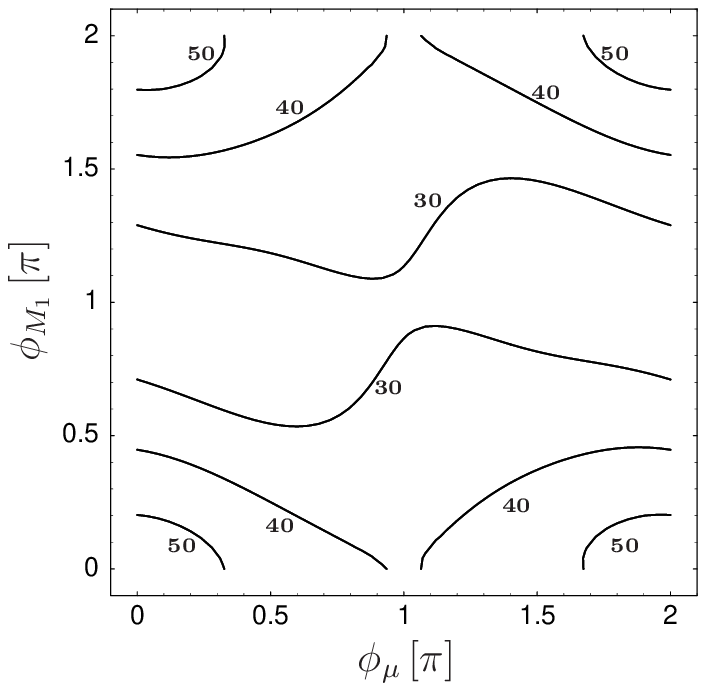,
height=31cm,width=19.4cm}}}
 \put(27,-70){\mbox{\epsfig{figure=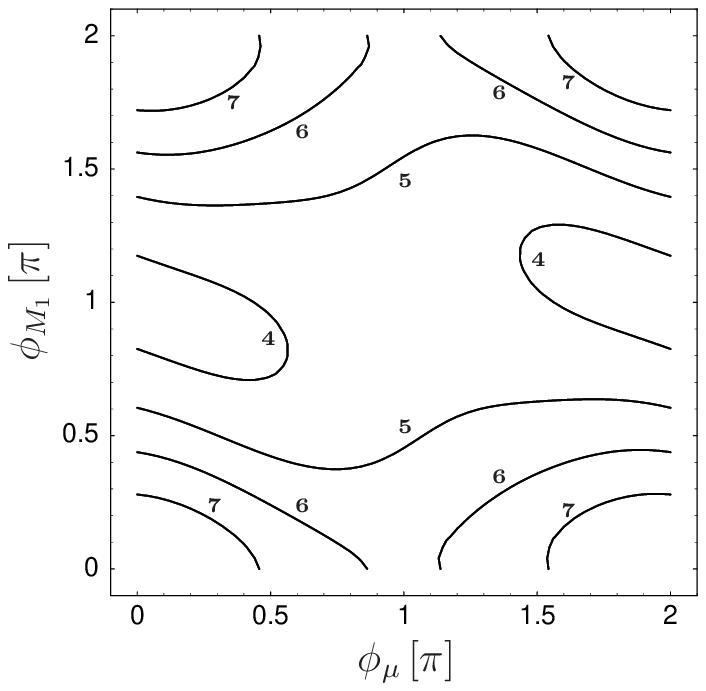,
height=31cm,width=19.4cm}}}
 \put(-53,-153){\mbox{\epsfig{figure=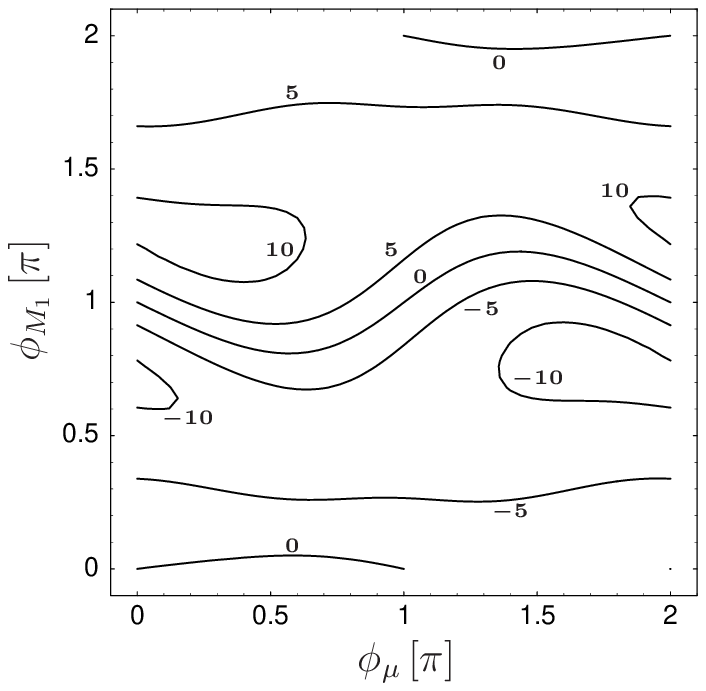,
height=31cm,width=19.4cm}}}
 \put(27,-153){\mbox{\epsfig{figure=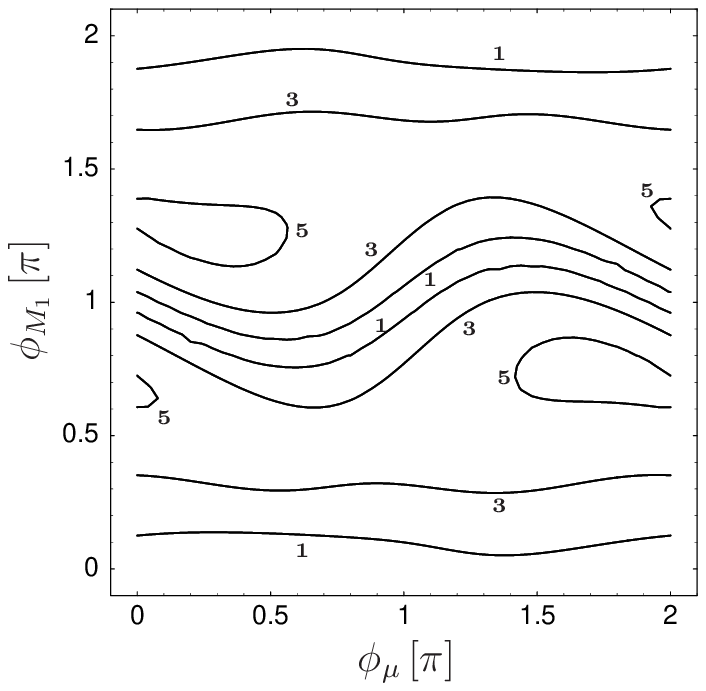,
height=31cm,width=19.4cm}}}
\put(8,207){\mbox{(a) $ \quad
\sigma(e^+e^- \to \tilde{\chi}^0_1 \tilde{\chi}^0_3) \,[\mbox{fb}] $}}
\put(88,207){\mbox{(b) $ \quad
\sigma(e^+e^- \to \tilde{\chi}^0_1 \tilde{\chi}^0_1 b \bar{b}) \,
[\mbox{fb}] $}}
\put(8,123){\mbox{(c) $ \qquad \qquad {\mathcal A}^{\rm T}_b \, [\%] $}}
\put(88,123){\mbox{(d) $ \qquad S^{\rm T}_b = 
|{\mathcal A}^{\rm T}_b|\sqrt{\sigma \cdot {\mathcal L}}$}}
\end{picture}
\end{center}
\vspace{-5cm}
\caption{
    Neutralino production
    $e^+e^- \to \tilde{\chi}^0_1 \tilde{\chi}^0_3$
    and decay
    $\tilde{\chi}^0_3 \to \tilde{\chi}^0_1 Z^0$, $Z^0\to b\bar{b}$
    at $\sqrt s=500$~GeV with 
    transverse beam polarizations 
    ${\mathcal P}_T(e^-)=0.9$, ${\mathcal P}_T(e^+)=0.6$, for
    Scenario~B, given in Table~\ref{tab1}.
    Contour lines in the $\phi_\mu$--$\phi_{M_1}$ plane for
    (a)~the production cross section
    $\sigma(e^+e^- \to \tilde\chi^0_1 \tilde\chi^0_3)$, 
    (b)~the total cross section 
    $\sigma(e^+e^- \to \tilde{\chi}^0_1 \tilde{\chi}^0_1 b \bar{b})$,
    (c)~the CP asymmetry ${\mathcal A}^{\rm T}_b$, and
    (d)~the statistical significance $S^{\rm T}_b$,
    for an integrated luminosity ${\mathcal L}=500$~fb$^{-1}$.
}
\label{N1N3_II}
\end{figure}

\begin{figure}[p]
\setlength{\unitlength}{1mm}
\begin{center}
\begin{picture}(150,200)
 \put(-53,-70){\mbox{\epsfig{figure=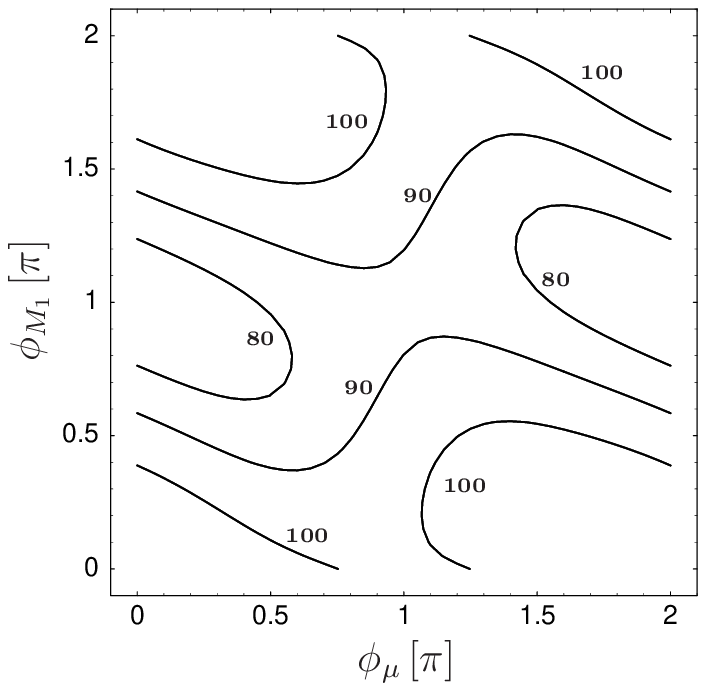,
height=31cm,width=19.4cm}}}
 \put(27,-70){\mbox{\epsfig{figure=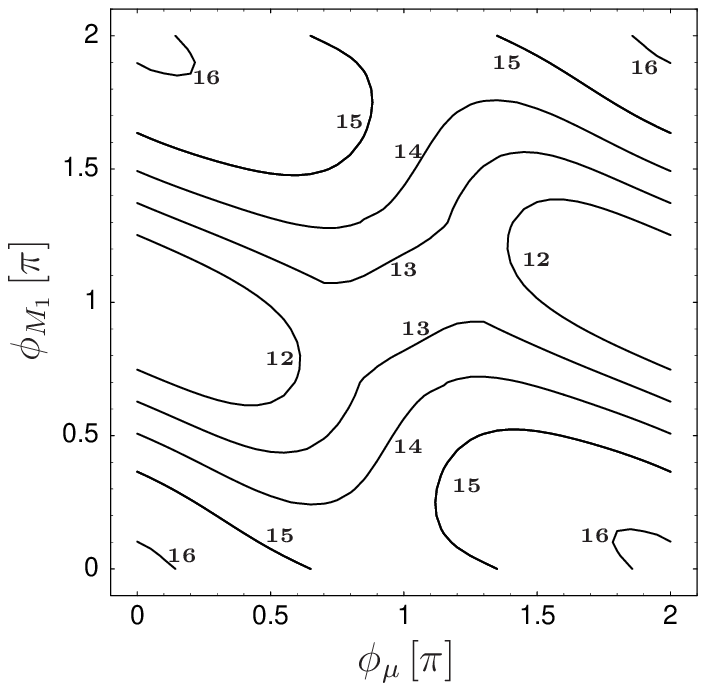,
height=31cm,width=19.4cm}}}
 \put(-53,-153){\mbox{\epsfig{figure=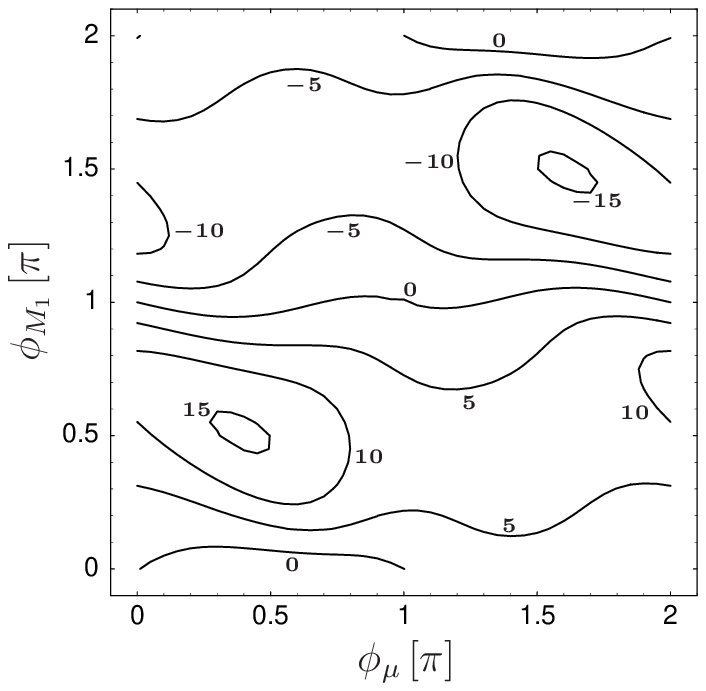,
height=31cm,width=19.4cm}}}
 \put(27,-153){\mbox{\epsfig{figure=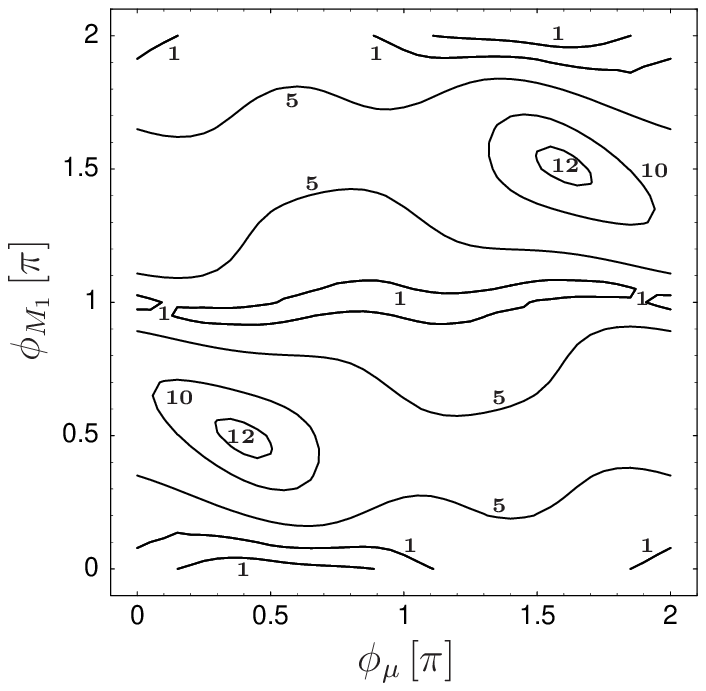,
height=31cm,width=19.4cm}}}
\put(8,207){\mbox{(a) $ \quad
\sigma(e^+e^- \to \tilde{\chi}^0_1 \tilde{\chi}^0_3) \,[\mbox{fb}] $}}
\put(88,207){\mbox{(b) $ \quad
\sigma(e^+e^- \to \tilde{\chi}^0_1 \tilde{\chi}^0_1 b \bar{b}) \,
[\mbox{fb}] $}}
\put(8,123){\mbox{(c) $ \qquad \qquad {\mathcal A}^{\rm L}_b \, [\%] $}}
\put(88,123){\mbox{(d) $ \qquad S^{\rm L}_b = 
|{\mathcal A}^{\rm L}_b|\sqrt{\sigma \cdot {\mathcal L}}$}}
\end{picture}
\end{center}
\vspace{-5cm}
\caption{
    Neutralino production
    $e^+e^- \to \tilde{\chi}^0_1 \tilde{\chi}^0_3$
    and decay
    $\tilde{\chi}^0_3 \to \tilde{\chi}^0_1 Z^0$, $Z^0\to b\bar{b}$
    at $\sqrt s=500$~GeV with 
    longitudinal beam polarizations 
    ${\mathcal P}_L(e^-)=0.9$, ${\mathcal P}_L(e^+)=-0.6$, for
    Scenario~B, given in Table~\ref{tab1}.
    Contour lines in the $\phi_\mu$--$\phi_{M_1}$ plane for
    (a)~the production cross section
    $\sigma(e^+e^- \to \tilde\chi^0_1 \tilde\chi^0_3)$, 
    (b)~the total cross section 
    $\sigma(e^+e^- \to \tilde{\chi}^0_1 \tilde{\chi}^0_1 b \bar{b})$, 
    (c)~the CP asymmetry ${\mathcal A}^{\rm L}_b$, and
    (d)~the statistical significance $S^{\rm L}_b$,
    for an integrated luminosity ${\mathcal L}=500$~fb$^{-1}$.
}
\label{N1N3_II_LongPol}
\end{figure}

In Fig.~\ref{N1N3_II_LongPol}, we show the phase dependence of
the production cross section
$\sigma(e^+e^- \to \tilde\chi^0_1 \tilde\chi^0_3)$,
the total cross section 
$\sigma(e^+e^- \to \tilde\chi^0_1 \tilde\chi^0_1 b \bar{b})$,
the asymmetry ${\mathcal A}^{\rm L}_b$
and the corresponding significance for longitudinally polarized beams 
$({\mathcal P}^L_{e^-},{\mathcal P}^L_{e^+})=(+0.9,-0.6)$ 
in scenario A of Table \ref{tab1}.
The asymmetry ${\mathcal A}^{\rm L}_b$ can reach $\pm15\%$,
the total cross section can be as large as 
$\sigma(e^+e^- \to \tilde\chi^0_1 \tilde\chi^0_1 b \bar{b})=16$~fb,
such that the significance can reach $ S^{\rm L}_b=12.$
In comparing these results, Fig.~\ref{N1N3_II_LongPol}, with
the results for transversely polarized beams, see Fig.~\ref{N1N3_II},
we find that the asymmetry ${\mathcal A}^{\rm L}_b$
can be measured with higher statistics compared to ${\mathcal A}^{\rm T}_b$. 
The reason is that the high degree of longitudinally polarized beams,
$({\mathcal P}^L_{e^-},{\mathcal P}^L_{e^+})=(0.9,-0.6)$,
enhances both the production cross section and the asymmetry,
whereas transversely polarized beams do not change the cross section,
see Fig.~\ref{N1N3_II}.

\medskip

\begin{figure}[t]
\setlength{\unitlength}{1mm}
\begin{center}
\begin{picture}(150,200)
 \put(-53,-70){\mbox{\epsfig{figure=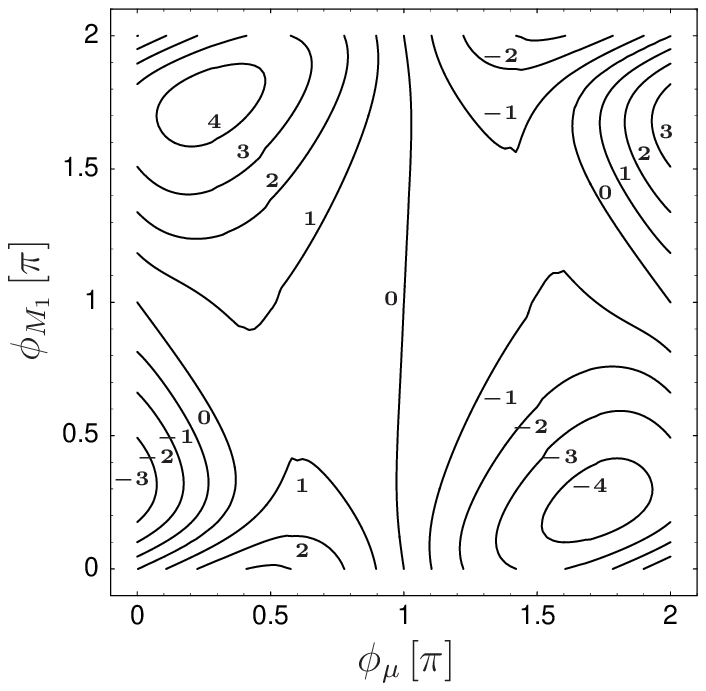,
height=31cm,width=19.4cm}}}
\put(27,-70){\mbox{\epsfig{figure=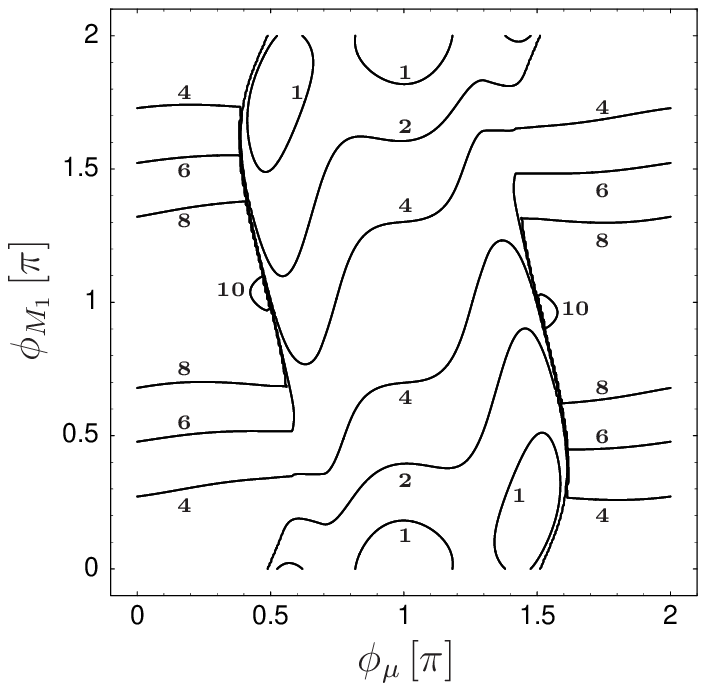,
height=31cm,width=19.4cm}}}
\put(8,207){\mbox{(a) $ \qquad \qquad {\mathcal A}^{\rm T}_b \, [\%] $}}
\put(88,207){\mbox{(b) $ \quad
\sigma(e^+e^- \to \tilde{\chi}^0_1 \tilde{\chi}^0_1 b \bar{b}) \,[{\rm fb}] $}}
\end{picture}
\end{center}
\vspace{-13.5cm}
\caption{
    Neutralino production
    $e^+e^- \to \tilde{\chi}^0_1 \tilde{\chi}^0_2$
    and decay
    $\tilde{\chi}^0_2 \to \tilde{\chi}^0_1 Z^0$, $Z^0\to b\bar{b}$
    at $\sqrt s=500$~GeV with transverse beam polarizations
    ${\mathcal P}_T(e^-)=0.9$ and ${\mathcal P}_T(e^+)=0.6$.
    Contour lines in the $\phi_\mu$--$\phi_{M_1}$ plane for
    (a)~the CP asymmetry ${\mathcal A}^{\rm T}_b$, and 
    (b)~the total cross section 
    $\sigma(e^+e^- \to \tilde{\chi}^0_1 \tilde{\chi}^0_1 b \bar{b})$,
    for Scenario~C, given in Table~\ref{tab1}.
}
\label{N1N2}
\end{figure}

\begin{figure}[t]
\setlength{\unitlength}{1mm}
\begin{center}
\begin{picture}(150,200)
 \put(-53,-70){\mbox{\epsfig{figure=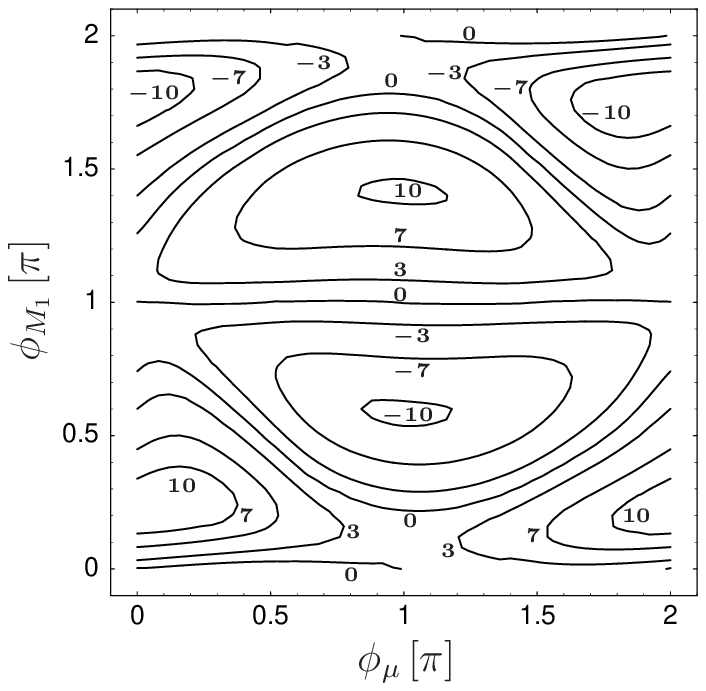,
height=31cm,width=19.4cm}}}
\put(27,-70){\mbox{\epsfig{figure=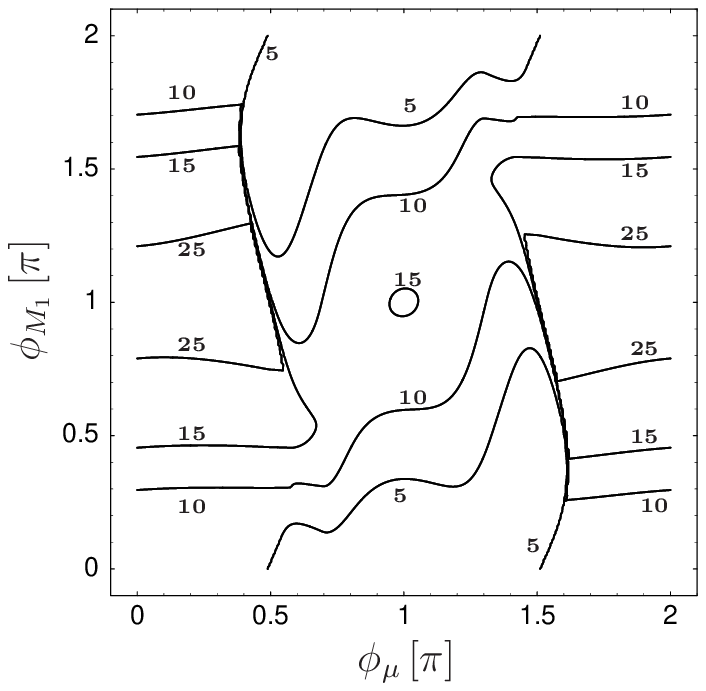,
height=31cm,width=19.4cm}}}
\put(8,207){\mbox{(a) $ \qquad \qquad {\mathcal A}^{\rm L}_b \, [\%] $}}
\put(88,207){\mbox{(b) $ \quad
\sigma(e^+e^- \to \tilde{\chi}^0_1 \tilde{\chi}^0_1 b \bar{b}) \,[{\rm fb}] $}}
\end{picture}
\end{center}
\vspace{-13.5cm}
\caption{
    Neutralino production
    $e^+e^- \to \tilde{\chi}^0_1 \tilde{\chi}^0_2$
    and decay
    $\tilde{\chi}^0_2 \to \tilde{\chi}^0_1 Z^0$, $Z^0\to b\bar{b}$
    at $\sqrt s=500$~GeV with longitudinal beam polarizations
    ${\mathcal P}_L(e^-)=-0.9$ and ${\mathcal P}_L(e^+)=0.6$.
    Contour lines in the $\phi_\mu$--$\phi_{M_1}$ plane for
    (a)~the CP asymmetry ${\mathcal A}^{\rm L}_b$, and 
    (b)~the total cross section 
    $\sigma(e^+e^- \to \tilde{\chi}^0_1 \tilde{\chi}^0_1 b \bar{b})$,
    for Scenario~C, given in Table~\ref{tab1}.
}
\label{N1N2_LongPol}
\end{figure}

\subsection{Neutralino $\tilde\chi^0_1 \tilde\chi^0_2$ production and decay
$\tilde\chi^0_2 \to \tilde\chi^0_1 Z^0 \to  \tilde\chi^0_1 b\bar{b}$}

In Fig.~\ref{N1N2}, we show the phase dependence of the 
CP asymmetry ${\mathcal A}^{\rm T}_b$, Eq.~(\ref{asym}), 
and the cross section for neutralino production 
$e^+e^- \to \tilde{\chi}^0_1 \tilde{\chi}^0_2$,
and decay
$\tilde{\chi}^0_2 \to \tilde{\chi}^0_1 Z^0$, $Z^0 \to b\bar{b}$,
for Scenario C, see Table~\ref{tab1}. 
The asymmetry reaches a maximum value of about $4.5\%$ for
$(\phi_{M_1},\phi_\mu) =(-\frac{\pi}{4},\frac{\pi}{4})$.
The production cross section
$\sigma(e^+e^- \to \tilde{\chi}^0_1 \tilde{\chi}^0_2)$ varies from
$19$~fb for $(\phi_{M_1},\phi_\mu) =(0,0)$, to $72$~fb
for $(\phi_{M_1},\phi_\mu)=(\pi,\pi)$. 
The branching ratio is maximal
${\rm BR}(\tilde\chi^0_2 \to  \tilde\chi^0_1 Z^0)=100\%$ for 
$\phi_\mu\lsim 0.4\pi$ and $\phi_\mu\gsim 1.6\pi$.
For $0.4\pi\gsim \phi_\mu\lsim 1.6\pi$, the $\tilde\chi^0_2$ decay channels  
$\tilde\chi^0_2\to\tilde\ell_R\ell$ are kinematical accessible, which leads 
to a strong $\phi_\mu$ dependence of the
cross section $\sigma(e^+e^- \to \tilde{\chi}^0_1 \tilde{\chi}^0_1 b \bar{b})$
around the threshold region.
In the region where ${\rm BR}(\tilde\chi^0_2 \to \tilde\chi^0_1 Z^0)=100\%$
($\phi_\mu\lsim 0.4\pi$ and $\phi_\mu\gsim 1.6\pi$) the total cross section
$\sigma(e^+e^- \to \tilde{\chi}^0_1 \tilde{\chi}^0_1 b \bar{b})$ varies 
from about $4$~fb to $10$~fb, see Fig.~\ref{N1N2}b. 
For Scenario C, with phases 
$(\phi_{M_1},\phi_\mu) =(-\frac{\pi}{4},\frac{\pi}{4})$,
the significance is $S_b\approx 2$, for 
${\mathcal L}=500$~fb$^{-1}$.

\medskip

In Fig.~\ref{N1N2_LongPol}, we show the asymmetry ${\mathcal A}^{\rm L}_b$
and the cross section for longitudinally polarized beams
$({\mathcal P}^L_{e^-},{\mathcal P}^L_{e^+})=(-0.9,+0.6)$ 
in scenario C of Table~\ref{tab1}. 
The asymmetry ${\mathcal A}^{\rm L}_b$ can reach $\pm10\%$,
and the total cross section can reach 
$\sigma(e^+e^- \to \tilde\chi^0_1 \tilde\chi^0_1 b \bar{b})=25$~fb.
Again the strong $\phi_\mu$ dependence of 
$\sigma(e^+e^- \to \tilde{\chi}^0_1 \tilde{\chi}^0_1 b \bar{b})$ 
in the regions where $\phi_\mu\sim 0.4\pi, 1.6\pi$ is because 
the decay channels $\tilde\chi^0_2\to\tilde\ell_R\ell$ are kinematical
accessible beyond this threshold. 
The significance to measure ${\mathcal A}^{\rm L}_b$ for $(\phi_{M_1},\phi_\mu)
=(-\frac{\pi}{4},\frac{\pi}{4})$ is $S^{\rm L}_b=7$ (${\mathcal L}=500$~fb$^{-1}$).
Once more we find that for longitudinally polarized beams, the asymmetry 
${\mathcal A}^{\rm L}_b$ can be measured with higher statistics than 
the asymmetry ${\mathcal A}^{\rm T}_b$.

\medskip

Finally, we have scanned the MSSM paramter space, 
and have found that in general 
the CP asymmetry ${\mathcal A}^{\rm T}_b$ is smaller for 
$\tilde\chi^0_1 \tilde\chi^0_2$ production
than for $\tilde\chi^0_1\tilde\chi^0_3$ production.
Note that in the parameter region where $M_2 > |\mu|$, for
which the largest values of ${\mathcal A}^{\rm T}_b$ are obtained for  
$\tilde\chi^0_1\tilde\chi^0_3$ production,
the two-body decay $\tilde\chi^0_2 \to  \tilde\chi^0_1 Z^0$
is kinematically forbidden.

\section{Summary and conclusions
        \label{Summary and conclusion}}

We have studied the impact of the CP violating MSSM phases
$\phi_{M_1}$ and $\phi_\mu$ on neutralino production 
$e^+e^- \to\tilde\chi^0_i \tilde\chi^0_j$
and decay $\tilde\chi^0_i \to \tilde\chi^0_n Z^0$, 
$Z^0 \to f \bar f$.
For transversely polarized beams, we have defined the
CP observable ${\mathcal A}^{\rm T}_f$, which is an asymmetry in the 
azimuthal distribution of the final fermions,
and is based on the triple product correlation Eq.~(\ref{eq:Toddcor}).
The CP-asymmetry depends bilinearly on the beam polarizations
${\mathcal A}^{\rm T}_{f} \propto {\cal P}_T^- {\cal P}_T^+$,
and receives CP sensitive contributions from spin-correlations
in the neutralino production process.
Due to the left-right structure 
of the $Z^0$ boson couplings to the final fermions,
the asymmtry is largest 
${\mathcal A}^{\rm T}_{b} = 6.3\times{\mathcal A}^{\rm T}_{\ell}$
for the hadronic decay $Z^0\to b\bar b$.

\medskip

In a numerical analysis for $\tilde\chi^0_1 \tilde\chi^0_2$
and  $\tilde\chi^0_1 \tilde\chi^0_3$ production, we have shown
that the asymmetry can be as large as
${\mathcal A}^{\rm T}_b=30\%$, for 
 ${\mathcal P}_T(e^-)=0.9$ and ${\mathcal P}_T(e^+)=0.6$.
The significance can be as large as 
$S^{\rm T}_b=5$, with  
an integrated luminosity ${\mathcal L}=500$~fb$^{-1}$.
We have compared the cross sections and asymmetries with
those which are accessible for longitudinally polarized beams.
For $\tilde\chi^0_1 \tilde\chi^0_2$ production, we have found
larger significances for the asymmetry 
${\mathcal A}^{\rm L}_f$ with longitudinally polarized beams.
Also for $\tilde\chi^0_1 \tilde\chi^0_3$ production
the asymmetry ${\mathcal A}^{\rm L}_f$ is in general
accessible with a larger statistical significance
compared to the asymmetry ${\mathcal A}^{\rm T}_f$ based
on transverse beam polarizations.
This is mainly because longitudinal beam polarizations
can greatly enhance the production cross section.
There are, however, parameter regions, e.g. for $M_2\gsim 300$~GeV
and $|\mu|\lsim 200$~GeV, where the asymmetry ${\mathcal A}^{\rm T}_f$
in $\tilde\chi^0_1 \tilde\chi^0_3$ production can be up to
about a factor five larger than the asymmetry ${\mathcal A}^{\rm L}_f$.

\medskip

We conclude that for neutralino 
$\tilde\chi^0_1 \tilde\chi^0_2$ production 
and subsequent two-body decay into the $Z^0$ boson,
longitudinally polarized beams give more statistics
than transversely polarized beams ($S^{\rm L}_f>S^{\rm T}_f$) to 
study the corresponding CP asymmetry 
in the neutralino sector of the MSSM.
However, for $\tilde\chi^0_1 \tilde\chi^0_3$
production, there are parameter regions where 
transversely polarized beams are advantageous since
$S^{\rm T}_f>S^{\rm L}_f$.
We emphasize that the asymmetries 
for longitudinally und transversely polarized beams are independent
of each other. They rely on different triple product correlations,
and depend on different interference channels in the neutralino
production process.
Therefore 
both options of polarized beams should be considered to
determine the phases $\phi_{M_1}$ and $\phi_\mu$ by measurements 
of the asymmetries ${\mathcal A}^{\rm T}_f$ and ${\mathcal A}^{\rm L}_f$
at the ILC.

\section{Acknowledgments}

This work is supported by the 'Fonds zur F\"orderung der
wissenschaftlichen Forschung' (FWF) of Austria, project. No. P18959-N16.
The authors acknowledge support from EU under the MRTN-CT-2006-035505
network programme.
This work was also supported by the SFB Transregio
33: The Dark Universe.

%\newpage
\vspace{1cm}

\begin{appendix}
        \noindent{\Large\bf Appendix}

\setcounter{equation}{0}
\renewcommand{\thesubsection}{\Alph{section}.\arabic{subsection}}
\renewcommand{\theequation}{\Alph{section}.\arabic{equation}}
\section{Momentum and polarization vectors
     \label{Momenta}}
\setcounter{equation}{0}

We introduce a coordinate frame in the laboratory system 
by choosing the $z$--axis along the momentum vector of 
the $e^-$ beam, and $x$ and $y$ according
to a right-handed coordinate system:  
\begin{eqnarray}
p^\mu_{e^-}=E_b(1,0,0,1)\quad\mbox{and}\quad 
p^\mu_{e^+}=E_b(1,0,0,-1)~, 
\end{eqnarray}
with the beam energy $E_b=\sqrt s/2$.
Then the 4-vectors of the transverse polarization
of the electron and positron beams are in the 
$x$--$y$ plane
\begin{equation}
t_{\pm}=(0,\cos\phi_{\pm},\sin\phi_\pm,0)~,
\label{eq:TPvec}
\end{equation}
where the azimuthal angles $\phi_+$ and $\phi_-$ describe
the orientation of the transverse beam polarizations,
which are fixed at any value $\phi_+,\phi_-\in [0,2\pi)$.
The four-momenta of the neutralinos are
\begin{eqnarray}
\label{eq:momentumneut}
p_{\chi_j}^{\mu}&=&q
(\frac{E_{\chi_j}}{q},\cos\phi \sin\theta,\sin\phi 
\sin\theta,\cos\theta)~,\nonumber\\
p_{\chi_i}^{\mu}&=&q
(\frac{E_{\chi_i}}{q},-\cos\phi \sin\theta,-\sin\phi \sin\theta,-\cos\theta)~,
\end{eqnarray}
where $\theta$ is the scattering angle and $\phi$ the azimuthal angle
of the production process.
The energies and momenta of the neutralinos are
\begin{equation}
\label{eq:energy}
E_{\chi_{i(j)}}=\frac{s+m^2_{\chi_{i(j)}}-m^2_{\chi_{j(i)}}}{2 \sqrt{s}}~,\qquad
q=\frac{\lambda^{\frac{1}{2}}(s,m^2_{\chi_i},m^2_{\chi_j})}{2 \sqrt{s}}~,
\end{equation}
with $\lambda(a,b,c)=a^2+b^2+c^2-2(a b + a c + b c)$.
The three spin-basis vectors $s^{b,\,\mu}_{\chi_j}$, $b=1,2,3$, of neutralino 
$\tilde{\chi}^0_j$ are 
\begin{eqnarray}
\label{eq:polvec}
s^{1,\,\mu}_{\chi_j}&=&
\left(0,\frac{{\vec{s}}^{\,\,2}_{\chi_j}\times{\vec{s}}^{\,\,3}_{\chi_j}}
{|{\vec{s}}^{\,\,2}_{\chi_j}\times{\vec{s}}^{\,\,3}_{\chi_j}|}\right)=
(0,-\cos\phi \cos\theta,-\sin\phi \cos\theta,\sin\theta)~,
\nonumber \\[3mm]
s^{2,\,\mu}_{\chi_j}&=&\left(0,
\frac{{\vec{p}}_{\chi_j}\times{\vec{p}}_{e^-}}
{|{\vec{p}}_{\chi_j}\times{\vec{p}}_{e^-}|}\right)=
(0,\sin\phi,-\cos\phi,0)~,
\nonumber \\[3mm]
s^{3,\,\mu}_{\chi_j}&=&\frac{1}{m_{\chi_j}}
\left(q, 
\frac{E_{\chi_j}}{q}{\vec{p}}_{\chi_j} \right)=
\frac{E_{\chi_j}}{m_{\chi_j}}
(\frac{q}{E_{\chi_j}},\cos\phi \sin\theta,\sin\phi \sin\theta,\cos\theta)~,
\end{eqnarray}
with $s_{\chi_j}^b\cdot p_{\chi_j}=0$ and $(s^{a}_{\chi_j}s^{b}_{\chi_j})=-\delta_{ab}$,
such that $\{\vec{s}^{\,\,1}_{\chi_j}, 
\vec{s}^{\,\,2}_{\chi_j}, 
\vec{s}^{\,\,3}_{\chi_j}\}$ build a right-handed system.

For the two-body decay $\tilde\chi_j^0\to\tilde\chi_n^0Z$,
the decay angle 
$\theta_{1} \angle (\vec p_{\chi_j},\vec p_{Z})$
is constrained  by $\sin\theta^{\rm max}_{1}= q^0/q$
for $q>q^0$,
where $q^0=\lambda^{\frac{1}{2}}(m^2_{\chi_j},m^2_Z,m^2_{\chi_n})/(2m_Z)$
is the neutralino momentum if the $Z^0$ boson is produced at rest.
In this case there are two solutions~\cite{Bartl:2003ck,Bartl:2004ut,Kittel:2004rp} 
\begin{eqnarray}
| \vec p^{\,\,\pm}_Z|= \frac{
(m^2_{\chi_j}+m^2_Z-m^2_{\chi_n})q\cos\theta_{1}\pm
E_{\chi_j}\sqrt{\lambda(m^2_{\chi_j},m^2_Z,m^2_{\chi_n})-
         4q^2~m^2_Z~(1-\cos^2\theta_{1})}}
        {2q^2 (1-\cos^2\theta_{1})+2 m^2_{\chi_j}}~.
\end{eqnarray}
If $q^0>q$, the decay angle $\theta_{1}$ is not 
constrained and there is only the physical solution 
$ |\vec p^{\,\,+}_Z|$. The momenta in the laboratory system are
 \begin{eqnarray}
&&   p_{Z}^{\pm} = (                        E_{Z}^{\pm},
            |\vec p_{Z}^{\,\,\pm}| \sin \theta_{1} \cos \phi_{1},
            |\vec p_{Z}^{\,\,\pm}| \sin \theta_{1} \sin \phi_{1},
            |\vec p_{Z}^{\,\,\pm}| \cos \theta_{1})~, \\[2mm]
&& p_{\bar f} = (                        E_{\bar f},
            |\vec p_{\bar f}| \sin \theta_{2} \cos \phi_{2},
            |\vec p_{\bar f}| \sin \theta_{2} \sin \phi_{2},
                                |\vec p_{\bar f}| \cos \theta_{2})~,\\[2mm]
&& E_{\bar f} = |\vec p_{\bar f}| = 
\frac{m_Z^2}{2(E_{Z}^{\pm}-|\vec p_{Z}^{\,\,\pm}|\cos\theta_{D})}~,
 \end{eqnarray}
with $\theta_{2} \angle (\vec p_{\chi_j},\vec p_{\bar f})$
and the decay angle 
$\theta_{D} \angle (\vec p_{Z},\vec p_{\bar f})$ given by
 \begin{equation}
\cos\theta_{D}=\cos\theta_{1}\cos\theta_{2}+
\sin\theta_{1}\sin\theta_{2}\cos(\phi_{2}-\phi_{1})~. 
  \end{equation}

\section{Phase space
     \label{Phase space}}
\setcounter{equation}{0}

The Lorentz invariant phase space element for the neutralino 
production (\ref{production}) and the decay 
chain (\ref{decay_1A})-(\ref{decay_2B}) can be decomposed
into the two-body  phase space 
elements~\cite{Bartl:2003ck,Bartl:2004ut,Kittel:2004rp}
\begin{eqnarray}
 &&d{\rm Lips}(s,p_{\chi_i },p_{\chi_n},p_{f},p_{\bar f}) =
         \nonumber \\ [2mm]
&&\frac{1}{(2\pi)^2}~d{\rm Lips}(s,p_{\chi_i},p_{\chi_j} )
~d s_{\chi_j} ~\sum_{\pm}d{\rm Lips}(s_{\chi_j},p_{\chi_n},p_{Z}^{\pm})
 ~d s_{Z}~d{\rm Lips}(s_{Z},p_{f},p_{\bar f})~,\label{Lips}
 \end{eqnarray}
\begin{eqnarray}
        d{\rm Lips}(s,p_{\chi_i },p_{\chi_j })&=&
        \frac{q}{4(2\pi)^2\sqrt{s}}~ d\Omega~, \\
        d{\rm Lips}(s_{\chi_j},p_{\chi_n},p_Z^{\pm})&=&
\frac{1}{2(2\pi)^2}~
\frac{|\vec p_Z^{\,\,\pm}|^2}{2|E_Z^{\pm}~q\cos\theta_1-
        E_{\chi_j}~|\vec p^{\,\,\pm}_Z||}~d\Omega_1~,\\
        d{\rm Lips}(s_{Z},p_{f},p_{\bar f})&=&
\frac{1}{2(2\pi)^2}~\frac{|\vec p_{\bar f}|^2}{m_Z^2}
        ~d\Omega_2~,
\end{eqnarray}
with $s_{\chi_j}=p^2_{\chi_j}$, $s_{Z}=p^2_{Z}$ and 
$ d\Omega_i=\sin\theta_i~ d\theta_i~ d\phi_i$.
We use the narrow width approximation for the propagators
\begin{equation}
\int|\Delta(\tilde\chi^0_j)|^2 $ $ d s_{\chi_j} = 
\frac{\pi}{m_{\chi_j}\Gamma_{\chi_j}}~, \quad \quad
\int|\Delta(Z^0)|^2 d s_{Z} = 
\frac{\pi}{m_{Z}\Gamma_{Z}}~.
\label{narrowwidth}
\end{equation}
The approximation is justified for
$\Gamma_{\chi_j}/m_{\chi_j}\ll1$,
which holds in our case with 
$\Gamma_{\chi_j}\lsim$ ${\mathcal O}(1 {\rm GeV}) $.
Note, however, that the naive 
${\mathcal O}(\Gamma/m)$-expectation of the error can easily receive
large off-shell corrections of an order of magnitude and more, 
in particular at threshold or due to interferences
with other resonant or non-resonant processes.
For a recent discussion of these issues, see~\cite{Hagiwara:2005wg,Berdine:2007uv}  

\end{appendix}

\end{document}